\documentclass[twocolumn]{bmcart}

\usepackage{amsthm,amsmath}
\usepackage[utf8]{inputenc} 
\usepackage{algorithm}
\usepackage{algorithmic}
\usepackage{mathtools}
\usepackage{amsfonts}
\usepackage{amssymb}
\usepackage{cite}
\usepackage{enumitem}
\usepackage{bigstrut}
\usepackage{subcaption}

\def\Jay{\textsc{Jay}}
\def\emphsf#1{{\small \textsf{#1}}}

\newcommand{\Hosts} {\mathcal{H}}
\newcommand{\hE} {h_{\rm C}}
\newcommand{\hL} {h_{\rm L}}
\newcommand{\hS} {h_{\rm S}}

\newcommand{\ttime} {T}
\newcommand{\TC} {\ttime_{{\rm C}}}
\newcommand{\TE} {\hat{\ttime}_{{\rm C}}}
\newcommand{\TI} {\ttime_{{\rm I}}}
\newcommand{\TO} {\ttime_{{\rm O}}}

\newcommand{\ener} {E}
\newcommand{\EC} {\ener_{{\rm C}}}
\newcommand{\EI} {\ener_{{\rm I}}}
\newcommand{\EO} {\ener_{{\rm O}}}

\newcommand{\POWER} {P}
\newcommand{\PC} {\POWER_{{\rm C}}}
\newcommand{\PI} {\POWER_{{\rm I}}}
\newcommand{\PO} {\POWER_{{\rm O}}}
\newcommand{\PU} {\POWER_{{\rm U}}}
\newcommand{\PD} {\POWER_{{\rm D}}}

\newcommand{\BI} {B_{{\rm U}}}
\newcommand{\BO} {B_{{\rm D}}}

\usepackage{url}
\usepackage{pifont}

\usepackage{booktabs} 
\usepackage{graphics}
\usepackage{algorithm}
\usepackage{algorithmic}
\usepackage{amsmath, amsfonts}
\usepackage{mathtools}
\usepackage{dsfont}
\usepackage{ulem}
\usepackage[htt]{hyphenat}
\usepackage{subcaption}
\usepackage{hyperref}

\usepackage{xcolor}

\newcommand\trackchange[2] {#2}

\startlocaldefs
\endlocaldefs

\begin{document}

\begin{frontmatter}

\begin{fmbox}
\dochead{Research}


\title{Energy-Aware Adaptive Offloading of Soft Real-Time Jobs in Mobile Edge Clouds}


\author[
  addressref={aff1},                   
  corref={aff1},
  email={joaquim.silva@fc.up.pt}   
]{\inits{J.}\fnm{Joaquim} \snm{Silva}}
\author[
  addressref={aff1},
  corref={aff1},
  email={ebmarques@fc.up.pt}
]{\inits{E.R.B.}\fnm{Eduardo R.B.} \snm{Marques}}
\author[
  addressref={aff1},
  corref={aff1},
  email={lmlopes@fc.up.pt}
]{\inits{L.M.B.}\fnm{Luís M.B.} \snm{Lopes}}
\author[
  addressref={aff1},
  corref={aff1},
  email={fmsilva@fc.up.pt}
]{\inits{F.}\fnm{Fernando} \snm{Silva}}

\address[id=aff1]{
  \orgdiv{CRACS/INESC TEC \& Department of Computer Science},             
  \orgname{Faculty of Sciences, University of Porto},          
  \city{Porto},                              
  \cny{Portugal}                                    
}





\begin{abstractbox}

\begin{abstract} 

\noindent
We present a model for measuring the impact of offloading soft
real-time jobs over multi-tier cloud infrastructures. The jobs
originate in mobile devices and offloading strategies may choose to
execute them locally, in neighbouring devices, in cloudlets or in
infrastructure cloud servers. Within this specification, we put
forward several such offloading strategies characterised by their
differential use of the cloud tiers with the goal of optimizing
execution time and/or energy consumption.
We implement an instance of the model using \Jay{}, a software
framework for adaptive computation offloading in hybrid edge
clouds. The framework is modular and allows the model and the
offloading strategies to be seamlessly implemented while providing the
tools to make informed runtime offloading decisions based on system
feedback, namely through a built-in system profiler that gathers
runtime information such as workload, energy consumption and available
bandwidth for every participating device or server. 
%
The results show that offloading strategies sensitive to runtime
conditions can effectively and dynamically adjust their offloading
decisions to produce significant gains in terms of their target
optimization functions, namely, execution time, energy consumption and
fulfillment of job deadlines.

\end{abstract}




\end{abstractbox}
\end{fmbox}

\end{frontmatter}



\section{Introduction\label{s:intro}}

The last decade witnessed an impressive evolution in the storage
and processing capabilities of mobile devices. Besides traditional
processing cores, these microprocessors feature multiple GPU cores and
also so called neural cores optimized for machine learning
applications such as deep-learning and have
reached performance levels comparable to laptop and some desktop
analogs~\cite{mobile_processors}. 

Despite these advancements, some computational jobs are too demanding
for mobile devices. Mobile cloud computing~\cite{FERNANDO201384} has
traditionally tackled this problem by offloading computation and data
generated by mobile device applications to cloud infrastructures. This
move spares the battery in the devices and, in principle, speeds-up
computation as the high-availability, elastic, cloud infra-structures
can adapt to the computing and storage demands of the jobs spawned by
the devices.

This offloading is, however, not without problems. Many mobile
applications involve the processing of locally produced data (e.g.,
video) and uploading such large volumes of data to cloud
infra-structures is time consuming and may not even be feasible from a
QoS point of view due to the high communication latencies
involved. Also, from an energy point of view, offloading jobs and/or
data to cloud infrastructures is globally highly inefficient.

Mobile edge clouds~\cite{edgeclouds} and cloudlets~\cite{cloudlets},
on the other hand, try to harness the resources of local networks of
devices and/or small servers, using device-to-device communication
technologies such as Wifi and Wifi-Direct, at the edge to perform
demanding computational jobs taking advantage of data locality to
minimize latency and global energy consumption. In this approach, a
given job is offloaded to one mobile device or a cloudlet in the
network vicinity of the originating mobile device.

\begin{figure}[t!]  
\centering 
\includegraphics[width=0.9\columnwidth]{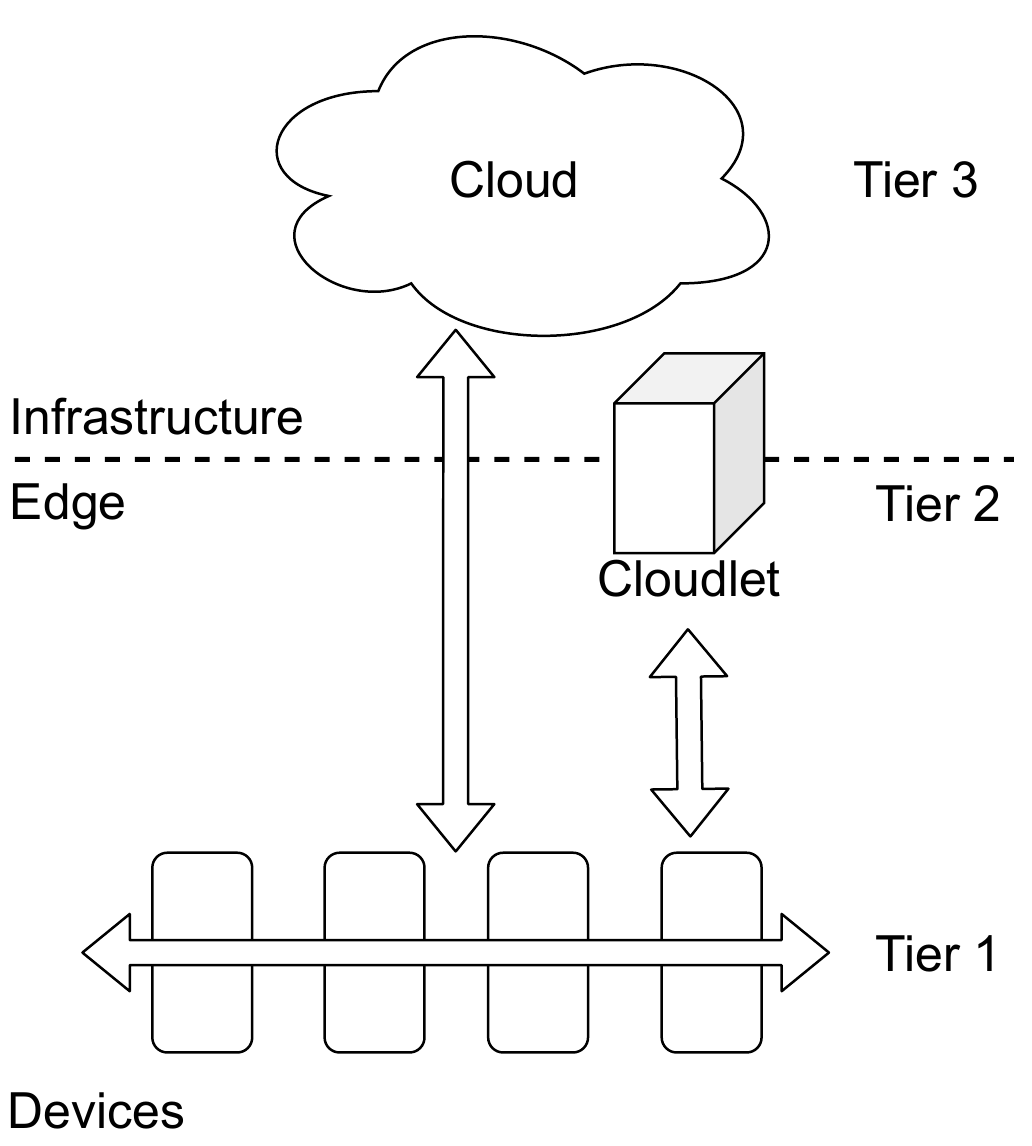} 
  \caption{A hybrid cloud environment.}  \label{fig:network}
\vspace{-0.2cm}
\end{figure}

The two approaches can be unified in a single, multi-tier architecture
(Fig.~\ref{fig:network}), with: (a) local networks of devices (Tier
1), with less capable processing but fast and privileged access to raw
data; (b) cloudlets directly accessible from the devices, with more
processing muscle and storage (Tier 2), and; (c) traditional
cloud infrastructures, featuring the highest performance and storage
resources (Tier 3).

Given this architecture and a mobile application that spawns
computational jobs, we consider the problem of offloading these jobs
over the tiers in such a way to optimize runtime metrics such as:
total execution time, global energy consumption, fulfillment of QoS
requirements. In general, the decision to offload (or not) a job is
supported by knowledge of observables as reported from the
participating devices and servers or inferred from data exchanges,
namely:  available network bandwidth, computational load at each
device and server, the battery status of the devices.

We previously introduced \Jay{}~\cite{fmec20_jay} as tool to
instantiate and experiment with different such cloud configurations,
offloading strategies and mobile applications. In that paper we
evaluated only latency-aware offloading strategies in several
cloud configurations, from mobile edge clouds formed by Android
devices up to 3-tier hybrid clouds, i.e., including also cloudlets and
infrastructure cloud server instances.

In this paper we put forward a unifying model for this architecture
upon which we can precisely specify the infra-structure parameters
(e.g., cloud tiers and topology), the application parameters (the rate and
size distribution of jobs, offloading strategy, job deadlines), the
observables (as described above) and the runtime metric function to
optimize. We then use~\Jay{} with the same object detection
application as in~\cite{fmec20_jay} but with different model instances
that include new QoS restrictions (jobs have deadlines) and different
optimization functions such as total execution time, per device energy
consumption and, total energy consumption.
%
The cloud configurations we experiment with in this paper do not
include tier-3 centralised cloud servers (e.g., Google Cloud, Amazon Web Services, or Microsoft Azure), as we would
not be able to directly measure vital runtime observables such as
energy consumption or at least infer them with enough confidence from
the underlying virtualisation infrastructure.
%


Thus, the main contributions of this paper are the following:
\begin{description}[style=unboxed,leftmargin=0cm]
  \item [1.] a model that specifies computational scenarios over hybrid
    edge/cloudlet/cloud topologies;
  \item [2.] a complete \Jay{} instance of the model that enables the execution of mobile applications over such network topologies through the definition of offloading strategies and optimization functions coupled with observables gathered at runtime, and;
  \item [3.] a case-study with an object detection application that
    generates jobs with deadlines while trying to optimize execution time, energy consumption or both.
\end{description}

\Jay{} and the model implementation presented here are available  at Github~\footnote{\url{https://github.com/jqmmes/Jay}}.

The remainder of this paper is structured as follows.
\trackchange{Related work appears right after the introduction.}{Related work is discussed in Section~\ref{s:rwork}.}
Section~\ref{s:model} provides a description of the model we use to
describe the aforementioned hybrid architecture and the computations
therein. Section~\ref{s:jay} describes the \Jay{}
framework. Section~\ref{s:setup} presents the the scenarios we model
in this paper and the experimental setup. Section~\ref{s:eval}
presents the results from the experiments and discusses their
implications.  Finally, Section~\ref{s:conclusion} ends the
paper with concluding remarks and a discussion of future work.

\section{Related work}\label{s:rwork}



\trackchange{Added reference to a few survey papers, including~\cite{shakarami2020survey} suggested by Reviewer 1}{
The general problem of computation offloading in mobile edge clouds received considerable
attention in the last two decades, as documented in recent surveys~\cite{kumar2013survey,mach2017mobile,shakarami2020survey}.
Our discussion of related work focuses on software
}
systems that, like \Jay{}, conduct adaptive offloading, and in
particular those \trackchange{typo}{that} implement energy-aware offloading policies.

A number of systems focuses on semi-automated offloading
to an edge cloud or centralised cloud infrastructure, without collaborative offloading between
mobile devices, in line with the mobile cloud computing paradigm~\cite{FERNANDO201384}.
In some systems of this kind,
e.g., Cuckoo~\cite{Kemp2012} or COSMOS~\cite{Shi2014}, offloading policies merely seek to minimize latency
without any energy awareness, and gains in energy consumption at the mobile device level
are at most a by-product of offloading computation.
A number of other systems support energy-aware offloading strategies supported by runtime profiling,
like AIOLOS~\cite{Verbelen2012},
MAUI~\cite{Cuervoy2010}, Phone2Cloud~\cite{Xia2014}, ThinkAir~\cite{Kosta2012}, or ULOOF~\cite{Neto2018}.

In the system model of all the former systems, energy consumption accounts only for the (local) mobile device
that hosts applications, typically the energy consumed in network transmission during offloading, rather than also the upper processing tiers in network and computation terms like \Jay{}, which may for instance also be battery-constrained
(e.g.,~\cite{fmec20_ramble,Rodrigues2018}) and in any case may have restrictions regarding energy
consumption (e.g., monetary costs).

In any case, the offloading policies of the systems still reflect
a concern for energy consumption, possibly in conjunction with latency:
AIOLOS allows one of two configurable policies that either optimize for latency or energy;
MAIUI minimizes energy consumption subject to a latency threshold constraint;
Phone2Cloud offloads jobs whenever the estimated latency for local execution exceeds a configurable
threshold, or when it perceive lower  energy consumption by the mobile device is attainable;
ThinkAir implements offloading policies that can seek to minimize only one of latency or energy consumption,
both latency and energy consumption (offloading must pay off in both dimensions compared to local execution),
and also optionally constrained by monetary posts due to the use of cloud services and; finally,
ULOOF evaluates local and remote execution cost functions that are parametrised by
a weight factor that can be used to attain a balance between latency and energy consumption.

Other types of systems enable collaborative offloading among mobile devices forming an edge cloud, and, in some cases,
also upper cloud tiers.
There are
systems of this kind which merely strive to optimize latency
like FemtoClouds~\cite{Habak2015}, Honeybee~\cite{honeybee}, Oregano~\cite{Sanches2020}, and P3-Mobile~\cite{p3mobile},
while others are explicitly energy-aware in diverse manners, \trackchange{}{discussed next}.


CWC~\cite{cwc} is a system for volunteer computing, where jobs are disseminated to a pool of mobile devices.
To prevent battery consumption and intrusive user experience,
jobs execute only when the devices are charging their batteries and have light computational loads,
and may also be paused to minimize battery charging times.

mClouds~\cite{mclouds} works over hybrid edge clouds that, like \Jay{}, may be composed of mobile devices in a ad-hoc wireless network, cloudlet and public clouds, offloading jobs to each of the tiers according
to connectivity conditions, and also (when multiple choices are available) by
a cost model with weights that balances execution time and energy consumption in a configurable manner.

MDC~\cite{mdc} is a system for collaborative offloading among mobile devices that
seeks to maximize the battery lifetime of the set of involved devices
by balancing energy consumption among them; this concern could provide an interesting refinement
to the HYBRID strategy in this paper, e.g., by factoring in battery levels of devices in addition
to their battery efficiency.

RAMOS~\cite{Gedawy2020} offloads jobs
over an edge cloud formed by heterogeneous mobile and IoT devices that act
as job workers, a concept borrowed from FemtoClouds~\cite{Habak2015}.
As in \Jay{}, the RAMOS scheduler can be parametrised to minimize job latency or
energy consumption, jobs have deadlines and are also executed in FIFO order by workers.
RAMOS' architecture is centralised, however. Jobs originate
and are scheduled in batches exclusively by a centralised controller node
in contrast to \Jay{}'s distributed architecture.

Synergy~\cite{Kharbanda2012} considers collaborative offloading
between devices in a peer-to-peer ad-hoc network, and, in order to maximise the devices' battery lifetime,
balances latency and energy consumption  by partitioning jobs among devices
while at the same time scaling the devices CPU frequencies.

\trackchange{RW Table Description}{

\trackchange{RW Table}{

\newcommand{\Vmark}{\ding{51}}%
\newcommand{\Xmark}{\ding{55}}%

\begin{table*}[h!]
\caption{Comparison between offloading systems.\label{t:rwork}}
\label{tab:related_work}
\resizebox{\textwidth}{!}{%
\begin{tabular}{@{}lcccccc@{}}
{\bf System} & {\bf Cloud architecture} &	{\bf Time-aware} & {\bf Energy-aware} & {\bf Deadlines}
& {\bf Scheduler} & {\bf Granularity} \bigstrut \\ 
\hline \hline
Jay & Configurable & \Vmark & \Vmark & \Vmark &  Configurable &  Single-job \bigstrut \\ \hline
Cuckoo~\cite{Kemp2012} & MCC & \Xmark & \Xmark & \Xmark & Local & Single-job\bigstrut \\
COSMOS~\cite{Shi2014} & MCC & \Xmark & \Xmark & \Xmark & Local & Single-job\bigstrut \\
\hline
AIOLOS~\cite{Verbelen2012} & MCC & \Vmark & \Vmark & \Xmark & Local & Single-job\bigstrut \\
MAUI~\cite{Cuervoy2010} & MCC & \Vmark & \Vmark & \Xmark & Local & Single-job\bigstrut \\
Phone2Cloud~\cite{Xia2014} & MCC & \Vmark & \Vmark & \Xmark & Local & Single-job\bigstrut \\
ThinkAir~\cite{Kosta2012} & MCC & \Vmark & \Vmark & \Xmark & Local & Single-job\bigstrut \\
ULOOF~\cite{Neto2018}  & MCC & \Vmark & \Vmark & \Xmark & Local & Single-job\bigstrut  \\
\hline
Femtoclouds~\cite{Habak2015} & Femtocloud & \Vmark & \Xmark & \Xmark & Centralized & Multiple-job \bigstrut \\
Honeybee~\cite{honeybee} & MEC & \Vmark & \Xmark & \Vmark & Local & Single-job\bigstrut \\
Oregano~\cite{Sanches2020}  & MEC & \Xmark & \Xmark & \Xmark & Centralized & Multiple-job\bigstrut \\
P3-Mobile~\cite{p3mobile} & MEC & \Xmark & \Xmark & \Xmark & Centralized & Parallel jobs\bigstrut  \\
\hline
CWC~\cite{cwc}  & Femtocloud & \Vmark & \Vmark & \Xmark & Centralized &  Parallel jobs\bigstrut \\
MDC~\cite{mdc} & MEC & \Vmark & \Vmark & \Xmark  & Centralized & Parallel jobs\bigstrut \\
RAMOS~\cite{Gedawy2020} & Femtocloud & \Vmark & \Vmark & \Vmark & Centralized & Multiple-job\bigstrut \\
Synergy~\cite{Kharbanda2012} & MEC & \Vmark & \Vmark & \Vmark & Local & Single-job \bigstrut \\
\hline \hline
\end{tabular}
}
\end{table*}
}

Summarising the above discussion, Table~\ref{tab:related_work}
provides a comparative overview of \Jay{} and the other systems
mentioned.  The table does so first in terms of cloud architecture,
making a distinction between: mobile cloud computing (MCC) systems,
where devices offload jobs to a centralised cloud infrastructure;
mobile edge computing (MEC) systems, where there is collaborative
offloading among mobile devices; and Femtocloud systems, where a set
of mobile devices is used as a worker pool for jobs fired by an
external host. The table next indicates awareness to runtime
information regarding time and/or energy, and the support for job
deadlines in the system model.  The two remaining columns characterize
the scheduler component responsible for offloading decisions concerning
its location and to the granularity of those offloading
decisions in terms of how many jobs are accounted for at once. A
scheduler may either operate locally per device or run on a central
peer, and the granularity type distinguishes between single-job,
multiple-job, and parallel job offloading by the scheduler. The latter
is a special class of multiple-job offloading in which all jobs are
bound by some type of parallel computation that is inherent to the job
model.

The main distinctive trait of \Jay{} is that it is configurable in terms of
target cloud architecture and scheduler operation. Thanks to a simple and
flexible design, each of the peers in a \Jay{} system instance may act as a
scheduler, a worker, or both, as illustrated by the variety of evaluation
scenarios we put forward later in the paper, meaning that one can use \Jay{}
for offloading using an MCC, MEC, or Femtocloud architecture, with per-device
schedulers or a centralised one. The offloading strategies we
instantiate and evaluate in \Jay{} are partially illustrative of
comparable time and/or energy-aware approaches found in other systems.
However, \Jay{} is not bound to any particular approach since offloading
strategies are configurable, and there is a general design for monitoring
runtime state information. Time-awareness is also
reflected in \Jay{} by the support of job deadlines, a feature
supported by only a few other systems. Finally, \Jay{} only supports
single-job scheduling granularity, a characteristic that is more in line with
on-the-fly offloading of independent jobs, as seen in most systems
discussed. In contrast, multiple-job granularity is usually associated
with the use of a centralised scheduler, a Femtocloud architecture, or
a computation model that embodies parallelism.

}

\section{System Model\label{s:model}}

\begin{figure*}[h!]
\centering 
\includegraphics[width=0.8\textwidth]{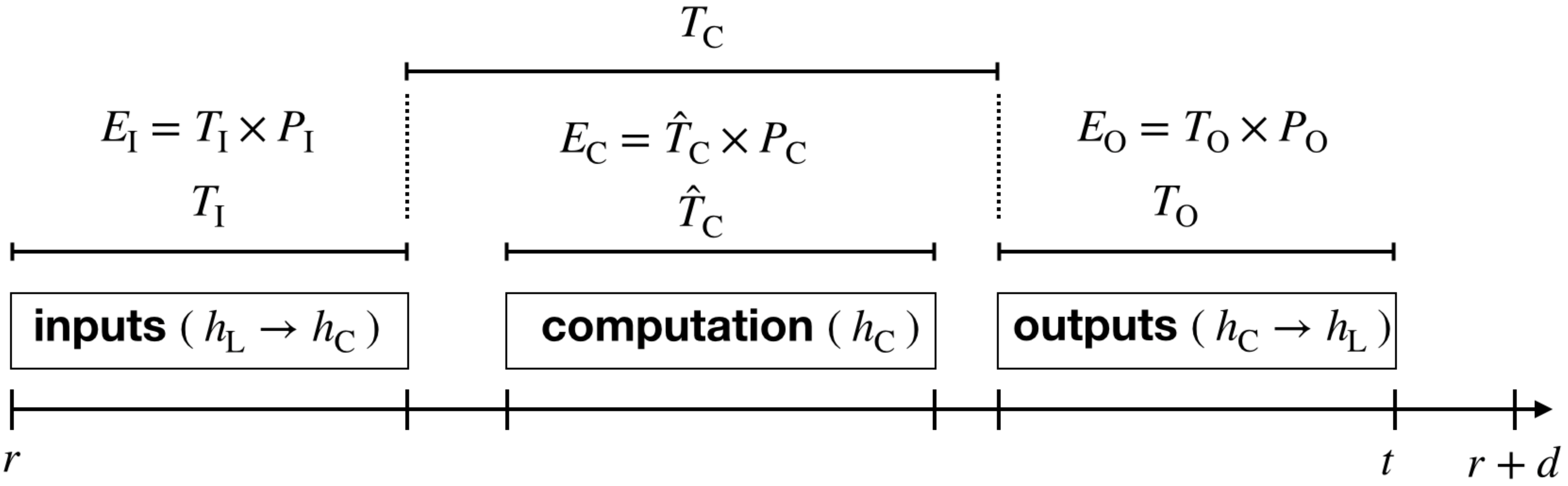}
\caption{Illustration of system model.\label{fig:model}}
\end{figure*}

\trackchange{Added subsection header for a clearer text organisation.}{
\subsection*{Overview}
}

We now put forward the system model for our adaptive offloading
framework.  

The overall rationale is as follows.  We consider a set of
hosts connected over a network, such that each host may generate
and/or execute soft real-time jobs over time.  A host may then execute
a mixture of local jobs and offloaded jobs on behalf of other hosts,
but it can also be that a host only generates or executes jobs.  In
this setting, offloading decisions can be informed and adaptive to
runtime conditions. Information broadcast amongst hosts regarding
variables such as network bandwidth and latency, host job load and
available energy provide the required feedback. 
We consider each job has a soft 
real-time nature, meaning that it has an associated relative
deadline expressing the maximum tolerable completion time for good
QoS, and also that it requires communication among hosts in the case
of offloading to supply job inputs (before the job's computation can
proceed) and obtain job outputs (when the computation is done).

\trackchange{}{
In what follows, we first lay out the base model 
concerning job characteristics and the time and energy costs
for offloading, and then present sample offloading strategies over that model.
} 

\trackchange{Added subsection header for a clearer text organisation.}{
\subsection*{Base definitions}
}

We associate each job~$j$ with a release time~$r(j)$, a termination
time~$t(j) > r(j)$, a relative deadline~$d(j)$, an originating
host~$\hL$ from a set of hosts~$\Hosts$, and, finally, a computation
host~$\hE$ also in~$\Hosts$.  When~$j$ is clear in context, these
properties are simply denoted respectively as~$r$,~$t$,~$d$, ~$\hL$,
and~$\hE$.  We say that the deadline of the job is fulfilled
if~$t \le r + d$.  Furthermore, we say the job executes locally
at~$\hL$ when~$\hL = \hE$, and that it is offloaded from~$\hL$
to~$\hE$ when~$\hL \neq \hE$.

In the scenario of runtime adaptive offloading, for job~$j$ and at
time~$r$, an offloading decision is made at time~$r$ to
determine~$\hE$. We assume that decision to be computed locally (at
$\hL$) and to have negligible overhead.  In the case of offloading
($\hL \neq \hE$), we assume that network communication needs to take
place between~$\hL$ and~$\hE$ for the inputs of~$j$ (data but possibly
also code) to be available at~$\hE$ before~$j$ starts, and, later,
once the computation~$j$ terminates, for the outputs to be transmitted
back from~$\hE$ to $\hL$.  This is illustrated in
Figure~\ref{fig:model}, along with the formulation for time and energy
overheads during offloading.

We consider the offloading decision to be informed by estimates of
completion time and energy consumption as follows.  Per each
host~$h \in \Hosts$ (including $\hL$) we model the estimated
completion time and energy consumption of a given job $j$, $\ttime(h)$ and $\ener(h)$,
respectively as:
$$
\ttime(h) = \TI(h) + \TC(h) + \TO(h)
$$
$$
\ener(h) = \EI(h) + \EC(h) + \EO(h)
$$
where time ($\ttime$) and energy ($\ener$) are factored into a sum of
three terms: $\TI$, $\EI$: the (time and energy) costs of input
offloading job~$j$; $\TC$, $\EC$: the (time and energy) costs of the
actual computation for job~$j$, and; $\TO$, $\EO$: the (time and
energy) costs of downloading outputs of job~$j$. Note that, given that
there is no need for network communication when~$h = \hL$, we should
necessarily have $\TI(\hL) = \EI(\hL) = \TO(\hL) = \EO(\hL) = 0$.

Regarding energy, our aim is not merely to account for the energy
consumption at the originating host ($\hL$) of a job, but also in the
computation host in the case of offloading ($\hE$).  This means first that
network I/O expressed by the~$\EI$ and~$\EO$ should account for
the energy consumption both in~$\hL$ and~$\hE$: sending inputs
from~$\hL$ to~$\hE$ requires energy to be consumed by $\hL$ in
uploading the inputs, and~$\hE$ to download, and vice-versa in the
case of outputs. Since~$\EI$ and~$\EO$ are respectively dependent on the transmission
times~$\TI$ and $\TO$ and the power consumption when doing so, we
model $\EI$ and $\EO$ as follows: 
$$
\EI(h) = \TI(h) \times \PI(h)
$$
$$
\EO(h) = \TO(h) \times \PO(h)
$$
where~$\PI$ and~$\PO$ are estimates for the power consumed per time unit
at both~$\hL$ and $\hE$\footnote{in these formulae, when clear from the context, the index of $h$ is
  omitted for the sake of simplicity}, 
when, respectively, sending job inputs from~$\hL$ to~$\hE$ and 
receiving job outputs at~$\hL$ from~$\hE$.

Moreover, the $\EC$ term reflects the cost of executing the job
remotely at~$\hE$, but, as illustrated in Figure~\ref{fig:model}, it
should only account for an estimate of the energy consumption
\emph{while}~$\hE$ is effectively performing the computation for
job~$j$ for an amount of time~$\TE$, rather than the energy
consumption over the total period~$\TC$, during which the job may at
times be pending (e.g., waiting for other jobs in $\hE$ to
complete). Thus, we write:
$$
\EC(h) = \TE(h) \times \PC(h)
$$
where~$\PC$ is an estimate for the power consumed per time unit at~$h$
due to the execution of the actual computation of~$j$ at~$h$.

\trackchange{Added subsection header for a clearer text organisation.}{
\subsection*{Offloading strategies}
}

\begin{figure*}[h!]
\centering 
\includegraphics[width=0.8\textwidth]{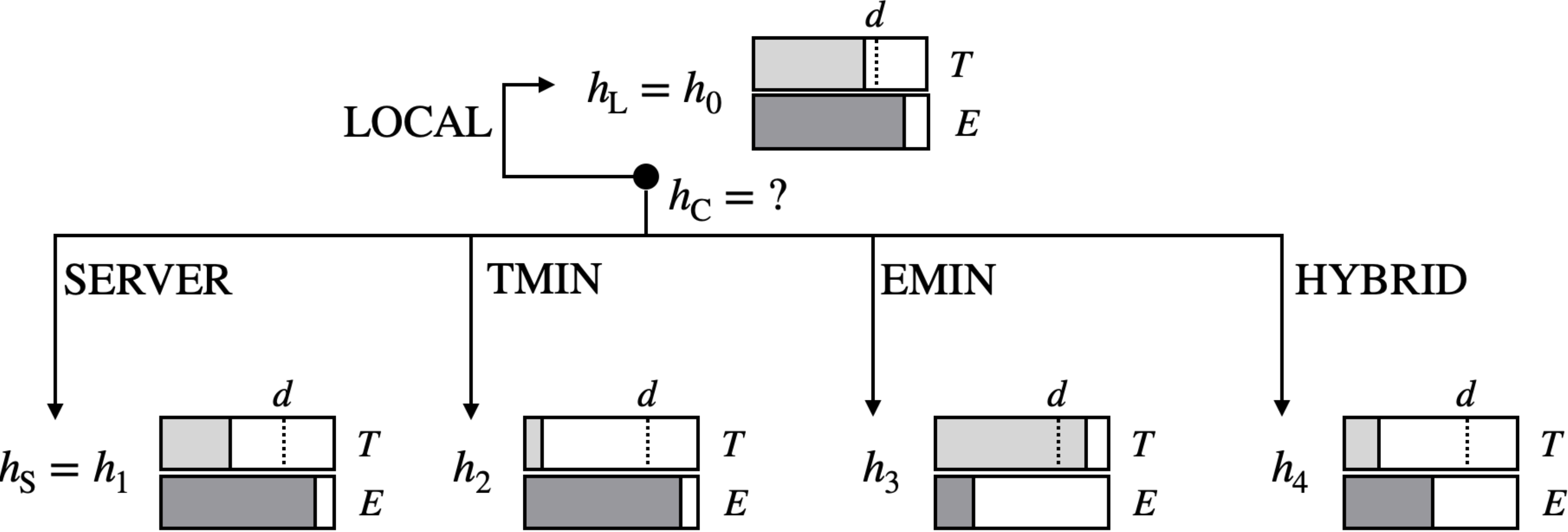}
\caption{Illustration of offloading strategies.\label{fig:strategies}}
\end{figure*}

\trackchange{}{We can now express offloading strategies that may take into consideration multiple metrics to decide where a computation will take place, e.g., completion time and energy consumption. We define several such strategies for which we present a thorough evaluation later in the paper.
Figure~\ref{fig:strategies} illustrates the rationale
in the offloading strategies. The example at stake concerns an offloading decision
for a job originating at host $\hL=h_0$ in an environment with four other hosts, 
$h_1$ to $h_4$, such that each host may have different values 
regarding the estimates for time and energy consumption ($\ttime$ and $\ener$).
}

The simplest case is that of no offloading, which we
designate as the {\rm LOCAL} strategy
\trackchange{}{
(i.e., as illustrated in Figure~\ref{fig:strategies}, jobs always execute locally).
}:
$$
{\rm LOCAL} \quad \equiv \quad \hE = \hL
$$
The choice may also be fixed to a special host $\hS \neq \hL$
(assuming $\hS$ does not generate jobs), for instance a cloud server
that is responsible for executing all jobs 
\trackchange{}{
(in Figure~\ref{fig:strategies}, $\hS$ is host $h_1$)}
, designated as the {\rm SERVER} strategy:
$$
{\rm SERVER} \quad \equiv \quad \hE = \hS
$$

\trackchange{}{
In the case illustrated in Figure~\ref{fig:strategies}, the {\rm LOCAL}
and {\rm SERVER} strategies would not lead to optimal choices with respect to time and/or energy, as there are hosts that can execute the job faster than $\hL$ and $\hS$ ($h_2$ and $h_4$), or that may consume less energy than $\hL$ and $\hS$ to do so ($h_3$ and $h_4$).
}
Adaptiveness comes into play if we account for the $\ttime$ and/or
$\ener$ estimates. 

 A strategy that seeks to minimize the completion
time of job, but ignoring energy consumption, can be defined as:
$$
{\rm TMIN} \quad \equiv \quad \hE = {\rm argmin}_{h \:\in\: \Hosts} \: \ttime(h)
$$
\trackchange{}{
Hence, in Figure~\ref{fig:strategies} we have $\hE = h_2$ for {\rm TMIN}.
}
In analogous manner, a strategy that seeks to minimize energy
consumption can be defined as:
$$
{\rm EMIN} \quad \equiv \quad \hE = {\rm argmin}_{h \:\in\: \Hosts} \: \ener(h)
$$ 
but it will not however attend to QoS requirements in terms of
deadline fulfilment, i.e., $\hE$ may be chosen regardless of whether
$\ttime(h_E) \le d$ or not. 
\trackchange{}{
This is illustrated in Figure~\ref{fig:strategies}, where $\hE = h_3$
for {\rm TMIN} but $T(h_3) > d$.
}
Additionally, the most energy-efficient
hosts will tend to be preferred. These hosts may possibly become
congested with too many jobs whose execution can therefore be much
delayed in time.  The above strategy can be refined meaningfully to
counter for these problems as:
$$
{\rm HYBRID} \quad \equiv \quad \hE = {\rm argmin}_{h \:\in\: \Hosts \::\: \ttime(h) \le d} \: \ener(h)
$$
balancing both time and energy costs and the fulfilment of $d$, as it
expresses that $\hE$ is chosen as the host which consumes less energy
amongst those that can satisfy the job deadline
($h \:\in\: \Hosts \::\: \ttime(h) \le d$). 
\trackchange{}{
This is illustrated in Figure~\ref{fig:strategies}, where $\hE = h_4$ is
the host with lower $\ener$ value, among those with a $\ttime$ value lower than $d$
(all except $h_3$).
}
 In the case where {\rm
  HYBRID} yields no result, i.e., no host is estimated to be able to
satisfy the job deadline, the offloading decision may for instance
fallback to {\rm TMIN} trying to complete the job as fast as possible
anyway or to simply cancel the job altogether.

\trackchange{Extended discussion}{

The  {\rm TMIN} and {\rm HYBRID} strategies may  lead to an imbalance between host loads, 
in the sense that most time-efficient and/or energy-efficient hosts will tend to have higher loads. This
may be counter-productive if the hosts are stake are battery-constrained and we wish for instance
to extend battery lifetime of all hosts as fairly as possible.
To spread the load more evenly, a balanced selection scheme of hosts amongst those
that can comply with a job deadline can be defined. For instance, a
balanced selection policy can be defined as:
$$
{\rm BALANCED} \quad \equiv \quad \hE = {\rm random} \:\{ h \:\in\: \Hosts \::\: \ttime(h) \le d \:\}
$$
i.e., $\hE$ is randomly selected among the hosts that can comply with the deadline.
In this case, the offloading choice may not be energy-optimal or time-optimal, but the random
choice will tend to promote a more balanced distribution of jobs. We could also refine ${\rm BALANCED}$ 
to be explicitly energy-aware by refining the definition with constraints for energy consumption or battery level thresholds in addition to the job's deadline. Also, in alternative, a round-robin job distribution could be considered
instead to enforce stricter load balancing.
}

\trackchange{}{Finally, we define a strategy that implements a form of ``restricted offloading'', a trait found in various systems discussed in Section~\ref{s:rwork}, such that jobs are only offloaded if local execution
is deemed unsuitable. That is, a job is only offloaded if the local host~$\hL$ is judged to be incapable of fulfilling the job's QoS, like deadlines in our case but also possibly other factors, e.g.,  those associated to network transmission in terms of energy, amount of data, or financial costs.
In line with this rationale, we formulate the ``local-first'' strategy ${\rm LF}[f]$, where~$f$ is the policy to apply in the case of offloading, as:
$$
{\rm LF} [f] \quad \equiv \quad \hE = \left\{
\begin{array}{@{}rl@{}}
\hL, & \mbox{ if } \ttime(\hL) \le d \bigstrut \\
f, & \mbox{otherwise}
\end{array}
\right.
$$
i.e., a job executes locally if the completion time estimate complies
with deadline, otherwise~$f$ is evaluated to decide where the job should run, e.g.,
we can define the ${\rm LF}[{\rm TMIN}]$, ${\rm LF}[{\rm HYBRID}]$, or ${\rm LF}[{\rm BALANCED}]$
strategies.

}

\section{The \Jay{} framework} \label{s:jay}


\Jay{} is a platform for the implementation and testing of
computation offloading strategies in hybrid clouds. \Jay{} is
provided as services implemented in Kotlin for Android OS, or as plain
Java Virtual Machines in other OSes (e.g., Linux or Windows). A
hybrid cloud may be composed of mobile devices, plus servers running
on cloudlets at the edge of the network or clouds accessible through
the Internet. \Jay{} instances in a hybrid cloud may host
applications that generate jobs and/or serve as computational
resources for offloading requests. Thus, the design makes no a priori
assumptions where applications reside, even if we are are particularly
interested in applications hosted on mobile devices.  In any case,
note that mobile devices can also serve offloading requests.
Furthermore, \Jay{}'s focus is not on data security/privacy preservation,
leaving this app-dependent.

\subsection*{Architecture}

\begin{figure*}[h!]
\includegraphics[width=0.8\textwidth]{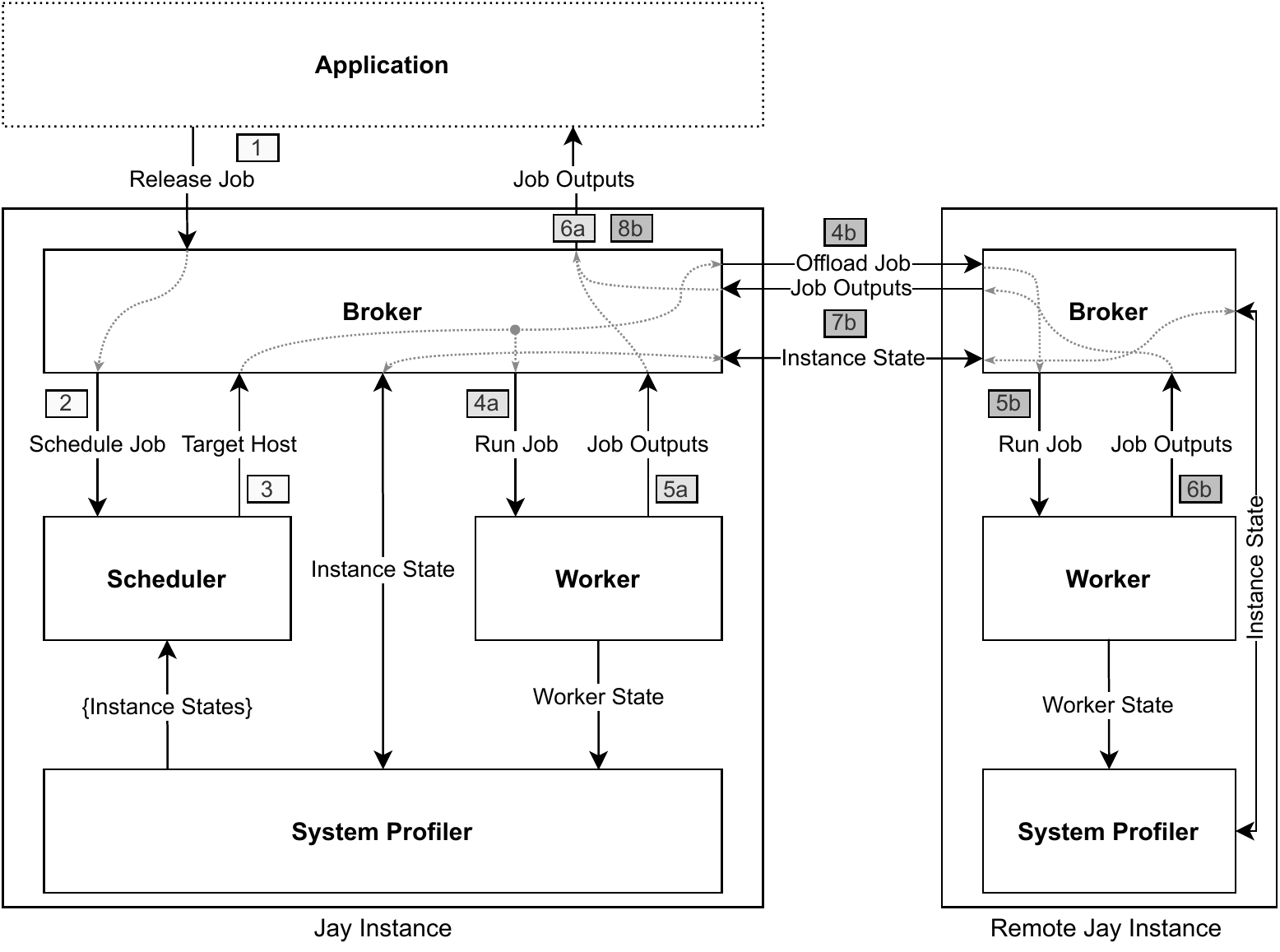}
\caption{Life cycle of a job in \Jay{}.}
\label{fig:job_lifecycle}
\end{figure*}

The architecture of \Jay{} is illustrated in Figure~\ref{fig:job_lifecycle}.
A single \Jay{} instance runs on a network host, comprising 4 services:
\emphsf{Broker}, \emphsf{Scheduler}, \emphsf{Worker} and \emphsf{System Profiler}.

The \emphsf{Broker} service mediates network interaction with external peers,
wrapping up any necessary requests and data interchange with the other
internal \Jay{} services on the same local instance. Local applications interact with the broker
for job execution, and broker-to-broker interaction occurs between \Jay{} instances
for job offloading and dissemination of state information. 

The internal state of each \Jay{}
instance over time is maintained by the \emphsf{System Profiler} service, reflecting
for instance energy consumption, current job workload, and network transmissions. 
All instances
disseminate their state periodically, hence the profiler
is also aware of the (latest known) state of remote \Jay{} instances.
The system profiler (on each instance) is then able to construct a global snapshot 
of all \Jay{} instances in the network at any given time. 
The goal is to use this dynamic global snapshot to guide the
offloading decisions while adapting to evolving runtime conditions.

Jobs are dealt with by the \emphsf{Scheduler} and \emphsf{Worker}
services.  The scheduler is responsible for choosing the host where to
run a job submitted to the local instance by an application. In
particular, it implements the offloading strategy.
The scheduler's choice for assigning a job is taken from the set of
all hosts having an active worker service, and can be based on state
information as reported by the system profiler. Note that this set of
active workers may include the local instance if it has an active
worker service. Also, the local worker state is also observed by the
profiler and included in the construction of the global state snapshot.
The worker is in turn responsible for the actual execution of jobs,
regardless of whether they are local or incoming from other hosts
through offloading requests. \Jay{} instances running only one of the
scheduler or worker services merely act as a job execution clients or
servers, respectively. On the other hand, instances may employ
different implementations for the scheduler and/or worker.

\subsection*{Job lifecycle}

In line with the interplay between \Jay{} services just described, we can trace   
the lifecycle of a job in terms of the stages indicated in Figure~\ref{fig:job_lifecycle},
as follows:
\begin{description}[style=unboxed,leftmargin=0cm]
\item[Job release:] an application first releases the job by placing an execution
request to the broker service (step \emphsf{1}). 
For simplicity, we will only consider the case where
the application resides on the same host as the broker,
even if \Jay{}'s architecture does not place such a constraint and
other setups may be interesting from an application standpoint,
e.g., to accommodate for jobs fired by IoT devices.
\item[Offloading decision:] the job execution request is passed by the broker over to the local scheduler {(\emphsf{2})} to determine the host that should execute the job.
The scheduler's decision (\emphsf{3}) may be that either the job executes locally (the local host
was chosen) or needs to be offloaded to the target host.
\item[Job execution:] for local execution, the broker passes the job for execution to the local worker (\emphsf{4a}),
and when the job completes {(\emphsf{5a})} the job outputs are delivered to the application {(\emphsf{6a})}.
In the offloading case, the job is sent to the target host
(\emphsf{4b}) for execution (\emphsf{5b}) and will, at some point,
produce the job outputs (\emphsf{6b}) that are then returned back to
the originating host (\emphsf{7b}) and, finally, delivered back to the application (\emphsf{8b}).
\end{description}


\subsection*{System instantiation}

We now present a sample system instantiation of \Jay{}, later
evaluated in the paper. It is composed of: a worker based on a FIFO
job queue; a configurable scheduler that may implement any of the
offloading strategies discussed in our system model, and; a system
profiler that estimates time and energy consumption due to computation
and network transmission at the local instance and aggregates it with
the state information disseminated by remote instances. The result is
a global snapshot of the state of the system that can be used to make
adaptive offloading decisions. A summary of the model instantiation
and associated notation is given in Table~\ref{t:notation}.

\begin{table}[h!]
\caption{Summary of model instantiation.}\label{t:notation}
\begin{tabular}{@{}llc@{}}
\\
\toprule
\multicolumn{2}{@{}l}{State variables per host $h$} & Unit \bigstrut\\ \midrule
$n$ & number of jobs at~$h$ & - \bigstrut\\
$\TE$ & computational time of a job & s  \bigstrut\\
$\BI$ & bandwidth for uploading data (to $h$) & B $/$ s \bigstrut\\
$\BO$ & bandwidth for downloading data (from $h$) & B $/$ s\bigstrut \\
$\PC$ & power consumption during job computation & W \bigstrut\\ 
$\PD$ & power consumption during data download & W \bigstrut\\ 
$\PU$ & power consumption during data upload & W \bigstrut\\ 
\\
\multicolumn{2}{@{}l}{Time estimates per host $h$} & Unit \bigstrut\\ \midrule
$\TC(h)$ & $\TE(h) \times \left(n(h) + 1\right) $ & s \bigstrut\\
$\TI(h)$ & $\BI(h) \times |j|_{\rm I}$ & s \bigstrut\\
$\TO(h)$ & $\BO(h) \times |j|_{\rm O}$ & s\bigstrut\\
$\ttime(h)$ & $\TI(h) + \TC(h) + \TO(h)$ & s \bigstrut\\
\\
\multicolumn{2}{@{}l}{Energy estimates per host $h$} & Unit \bigstrut\\ \midrule 
$\EC(h)$ & $\TE(h) \times P_C(h)$ & W $\cdotp$ s\bigstrut\\
$\EI(h)$ & $\TI(h) \times \left( \PU(\hL) + \PD(h)\right)$ & W $\cdotp$ s\bigstrut \\
$\EO(h)$ & $\TO(h) \times \left( \PD(\hL) + \PU(h)\right)$ & W $\cdotp$ s \bigstrut\\
$\ener(h)$ & $\EI(h) + \EC(h) + \EO(h)$ & W $\cdotp$ s \bigstrut\\
\bottomrule
\end{tabular}
\end{table}

\subsubsection*{Profiler overview}

\begin{figure*}[h!]
\includegraphics[width=0.7\textwidth]{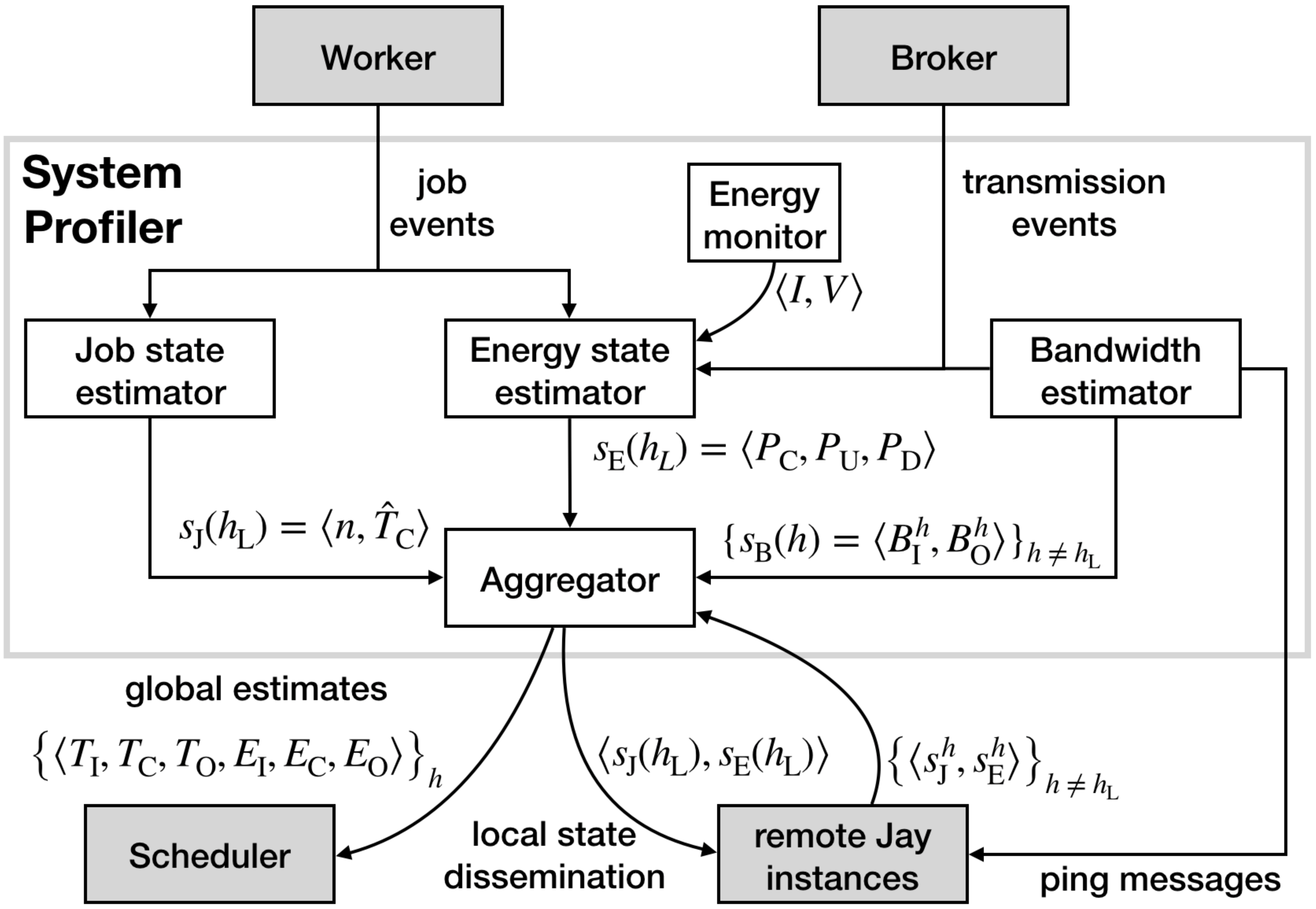}
\caption{State estimation by the system profiler.\label{fig:profiling}}
\end{figure*}

The profiler is responsible for the state estimation driving adaptive offloading, 
with the functionality illustrated in Figure~\ref{fig:profiling}.
The first aim of the profiler is to estimate and disseminate the state
of the local instance ($\hL$), and aggregate similar state reported by the
remote instances ($h\neq\hL$), as shown lower right in the figure. The
second aim is to use the state information for all available hosts
($\hL$ and other hosts $h\neq\hL$) to compute estimates for all hosts
for the time ($\TI$, $\TC$, and $\TO$) and energy ($\EI$, $\EC$, and
$\EO$) to run a job, and feed that information to the local scheduler,
as illustrated in the lower left portion of the figure.

The locally derived information comprises three components and their
estimators (modules) also shown in the figure: $s_{\rm J}$, the state
of local jobs, derived by the job state estimator; $s_{\rm E}$, the
energy consumption state, derived by the energy state estimator;
and~$s_{\rm B}$, the bandwidth for communication between~$\hL$ and
every other host, derived by the bandwidth estimator. To accomplish
their task, the estimators feed on notification events provided by the
local worker and local broker regarding job and transmission events
respectively, and runtime profiling of energy consumption and
bandwidth measurements. Note that only $s_{\rm J}$ and~$s_{\rm E}$
need to be disseminated among instances, whereas the~$s_{\rm B}$
information for all hosts is derived locally at each instance.

\subsubsection*{Job computation}

The worker executes jobs in order-of-arrival, one at a time, and
non-preemptively until completion. Pending jobs are kept on hold in a
FIFO queue. This scheme is not adaptive to deadlines or other job
characteristics. But, on the other hand, it allows for a simple
estimation of the termination time for a released job that is not
affected by the arrival of new jobs or more generally by overall
variations in the system workload.  Assume that job~$j_1$ starts
running at time $t$, that jobs $j_2, \ldots, j_n$ are queued, and that
there is an estimate~$\Delta_i$ for the time~$j_i$ takes to
execute. Then an estimate for the termination time of~$j_i$ is
simply~$t + \Delta_1 + \ldots + \Delta_i$.

Note that this ``stable'' estimate, derived from limited information,
would be impossible to achieve if we were to resort, for instance, to
an earliest-deadline first (EDF) scheme in preemptive or
non-preemptive form. In this case, the arrival of new jobs could
potentially invalidate a previous estimate made during an offloading
decision, and raise the need to model/estimate a worst-case behavior
for job arrivals.  

The worker interacts with the system profiler by supplying a
notification whenever a job is queued, starts, and ends. With this
information, the profiler can compute an estimate of the job
execution time and the worker's queue size and composition.
From these quantities we can in turn derive estimates
for~$\TC$~and~$\TE$ (cf. Figure~\ref{fig:model}).
Feeding on the local worker information, the
current profiler estimates~$\TE$ using a moving average of the
execution time of jobs, and $\TC$ as:
$$
\TC(h) = \left(n(h) +1\right) \times \TE(h)
$$ 
The formula above simply expresses that the time to execute a job~$j$
will have to account for the wait for up to~$n$ jobs to complete at
host~$h$, plus the time to actually execute~$j$. 
The estimate implicitly assumes however that job execution time 
tends to be uniform, i.e., there is only one class of job and their execution
is regular. This in the case of the jobs we consider for evaluation later,
but the scheme could be generalised, e.g. to handle several classes of jobs
by accounting for the number of jobs per class, and irregular jobs
by accounting for different job input sizes and/or considering execution
time percentiles rather than a plain moving average.

\subsubsection*{Network transmission times}

In order to estimate network transmission times, the profiler issues
periodic ping (round-trip) messages to all hosts in the network. The
information gathered from these messages allows the bandwidth
estimator at each host~$h$ to maintain a moving average for uploading
and downloading bandwidth measures, $\BI(h)$ and~$\BO(h)$,
respectively. These estimates are further refined with information
gathered from broker notifications regarding the observed bandwidths when jobs
(and their inputs) are uploaded to, or their outputs are downloaded from, a
remote \Jay{} instance. Assuming that the sizes of the inputs
($|j|_{\rm I}$) and outputs ($|j|_{\rm O}$) of a job $j$ are known,
the profiler estimates $\TI$ and $\TO$ as follows:
$$
\TI(h) = \BI(h) \times |j|_{\rm I}
$$
$$
\TO(h) = \BO(h) \times |j|_{\rm O}
$$

\subsubsection*{Power consumption estimates}

The energy state monitor is responsible for maintaining running
estimates for the power cost terms~$\PC$, ~$\PU$, and~$\PD$ that
correspond, respectively, to the power consumption per time unit when
executing jobs at, uploading data from, and downloading data to
the local host.  In contrast to other approaches, \Jay{} produces
estimates without resorting to any a priori, usually device-specific,
derived model for power consumption. As such, power consumption
estimates may be more crude but, on the other hand, reflect more
closely the energy dynamics of the system at any given moment.

At any given time, a \Jay{} instance may be idle, performing
computation, or transmitting data.  This can be inferred by listening
to job events from the worker and transmission events by the broker or
bandwidth monitor. Whenever the worker starts a job, the active job
computation status flag is enabled, meaning the ensuing energy
consumption should reflect on the $\PC$ estimate.  The same flag is
disabled whenever the job ends. The~$\PU$ and~$\PD$ estimates are
derived similarly, using upload and download status flags that are
enabled and disabled according to the start and end events for uploads
and downloads by the broker or the bandwidth monitor. The power
consumption estimates are updated as moving averages in accordance to
the values of the status flags, but when only one of the flags is active.
Power consumption measures are obtained in a device-specific
manner through an energy monitor. For instance, 
by measuring the current~$I$
and the voltage~$V$ in a device, 
a simple estimate for the power consumption would be~$P = I \times V$ (using Ohm's Law). 
With estimates for~$\PC$,~$\PU$, and~$\PD$ plus~$\TE$,~$\TI$, and~$\TO$ 
we can in turn express the corresponding energy costs for jobs
as follows:
$$
\EC(h) = \TE(h) \times P_C(h) 
$$
$$
\EI(h) = \TI(h) \times \left( \PU(\hL) + \PD(h)\right)
$$
$$
\EO(h) =\TO(h) \times \left( \PD(\hL) + \PU(h)\right)
$$
Note that, in line with the starting discussion for the system model,
$\EC$ depends on~$\TE$ (the effective computation time at~$h$) rather
than~$\TC$ (the entire time span the job is at~$h$), while~$\EI$
and~$\EO$ reflect the energy costs both at~$\hL$ and~$h$.

\section{Experimental Setup} \label{s:setup}

We used the \Jay{} framework to evaluate the algorithms presented in
Section~\ref{s:model}, namely: {\rm LOCAL}, {\rm SERVER}, {\rm
  TMIN} and {\rm HYBRID}. 

\subsection*{Devices}

\begin{table*}[t!]
\caption{Device characteristics.}
\label{tab:devices}
\centering
\begin{tabular}{@{}llllll@{}}
\toprule
Device & Year & CPU  & RAM & Battery & OS \bigstrut \\
\midrule
Cloudlet & 2015 & Intel i7-6700K, 4x4.0 GHz & 16 GB & N/A & Ubuntu 20.04 LTS\bigstrut \\
Google Nexus 9 & 2014 & Nvidia Tegra K1, 2x2.3 GHz & 2 GB& 6.7 Ah& Android 8.1 \bigstrut \\
Google Pixel 4 & 2019 & Snapdragon 855, 1x2.84/3x2.42/4x1.78 GHz & 6 GB & 2.8 Ah & Android 11 \bigstrut \\
Samsung Galaxy S7e & 2016 & Exynos 8890 Octa, 4x2.3/4x1.6 GHz & 4 GB & 3.6 Ah& Android 10\bigstrut \\
Samsung Galaxy Tab S5e & 2019 & Snapdragon 670, 2x2.0/6x1.7 GHz & 6 GB & 7.0 Ah& Android 10\bigstrut \\
Xiaomi Mi 9T & 2019 & Snapdragon 730, 2x2.2/6x1.8 GHz & 6 GB & 4.0 Ah & Android 10\bigstrut \\
\bottomrule
\end{tabular}%
\end{table*}

The experimental setup consisted of~5 Android devices and a PC-based
cloudlet. In experiments detailed in the next section, Android devices 
are used as job generators and executors, while the cloudlet is
used as job executor only.
Their characteristics are summarised in
Table~\ref{tab:devices}. It can be seen that the Android devices are
quite heterogeneous in terms of CPU, RAM and battery capacity, as well
as in terms of their Android OS version. Another important aspect is
that the cloudlet has significantly more RAM (16 GB) than all Android
devices and, as illustrated in detail later, also uses a higher
performance CPU configuration. All devices were connected to the same
local network, via an ASUS RT-AC56U router, featuring a 2.4~GHz
300~Mbit/s WiFi connection for the Android devices and a 1~Gb/s
Ethernet connection for the cloudlet.

Prior to each experiment, all Android devices had their batteries
charged by at least 50\%, to prevent interference from builtin power
saving mechanisms. They were then disconnected from the power outlet
using a smart plug controlled remotely by a script.  For monitoring
energy we used the standard Android {\tt BatteryManager}
API\footnote{\url{https://developer.android.com/reference/android/os/BatteryManager}}.
This API provides current intensity ($I$) and voltage ($V$)
information, respectively, through the {\tt
  BATTERY\_PROPERTY\_CURRENT\_NOW} counter and the {\tt
  EXTRA\_VOLTAGE} notifications. The API is available and reliable
across all Android versions and devices we tested, from which we
estimated the instantaneous power consumption ($P = I \times V$).  We
used this approach uniformly for all devices, even if in some
devices/Android versions the API provided a richer set of attributes such as
the {\tt BATTERY\_PROPERTY\_CURRENT\_AVERAGE}, for average current
intensity, and {\tt BATTERY\_PROPERTY\_ENERGY\_COUNTER}, for the
remaining battery power. As for the cloudlet, it had a permanent 220~V
power supply. Its power consumption was monitored using a Meross
MSS310 energy plug.



\subsection*{Benchmark Application}

\begin{figure}[h!]
\centering
\includegraphics[width=0.5\columnwidth]{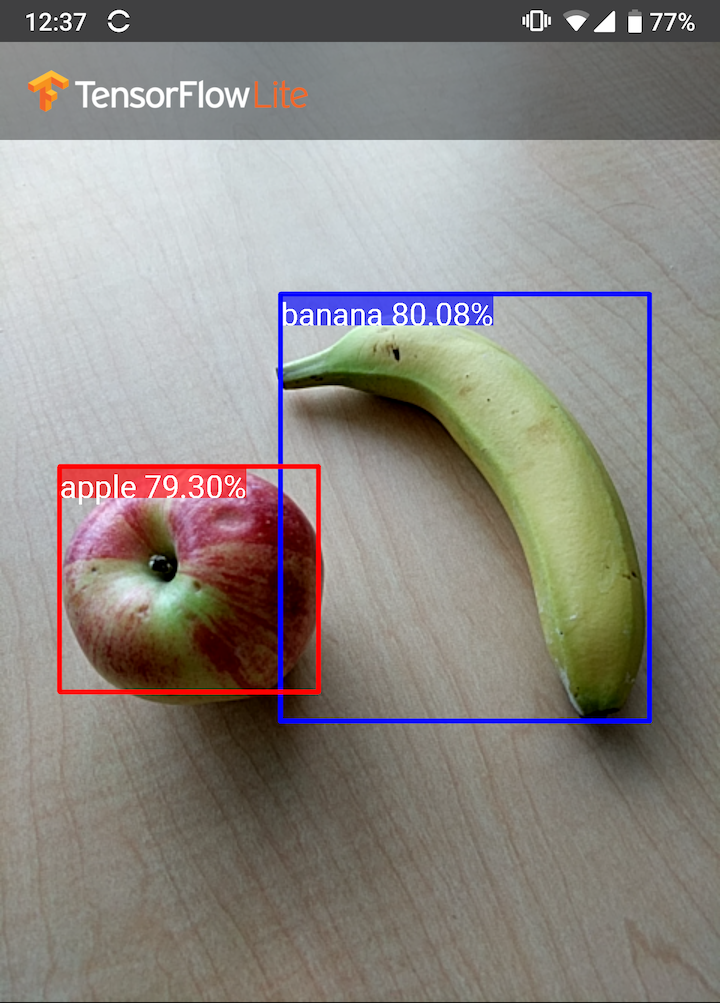}
\caption{Object detection in images.}
\label{fig:tfapp}
\end{figure}

We used a benchmark application that fires jobs for object detection
in images using deep learning, similar to one we employed in previous
work~\cite{fmec20_jay}.  As illustrated in Figure~\ref{fig:tfapp},
each object detection job takes an image as input and yields a set of
objects detected in the image along with corresponding bounding boxes
and confidence scores. Here, we used a ``headless'' variant of this
computational job with no GUI or human intervention. Overall, this
type of computation is increasingly common to classify static images
or live video frames in mobile devices~\cite{tomm17,www19dl}. It makes
for an interesting case-study for offloading since jobs can be
computationally intensive and may require high network bandwidth to
transfer images. This can happen if the number of spawned jobs is
large or if a QoS restriction is added to their execution (e.g., a
deadline), or both.

We make use of a MobileNet SSD model variant~\cite{mobilenet} trained
with COCO~\cite{coco}, a popular image dataset used for benchmarking
object detection models that contains 2.5 million labeled instances
for 80 object types in more than 300.000 images.  The specific model
we use is \texttt{ssd\_mobilenet\_v1\_fpn\_coco}, available in
standard TensorFlow (TF)~\cite{tensorflow} and TensorFlow Lite
(TFLite) format from TensorFlow's Object Detection
Zoo\footnote{\url{https://github.com/tensorflow/models/blob/master/research/object_detection/g3doc/tf2_detection_zoo.md}}.
The object detection job code, adapted from a TensorFlow
tutorial\footnote{\url{https://github.com/tensorflow/examples/tree/master/lite/examples/object_detection/android}},
has been incorporated into two distinct Kotlin modules, each linked
with the \Jay{} core library. One module is used in the cloudlet
(Linux) and employs the standard TF library. The other is used in the
Android devices and employs TFLite. Besides CPUs, TF and TFLite may
employ GPUs, if available. In the case of TFLite in Android, it can
also use the specialised Google Neural Networks
API~\footnote{\url{https://developer.android.com/ndk/guides/neuralnetworks/}}. Nevertheless,
we configured the device's Kotlin module to use only CPUs given that
this basic option works for all devices and operating system versions.


\begin{table*}[h!]
\caption{Baseline results per device.}
\label{tab:baseline}
\centering
\begin{tabular}{@{}l@{\quad}r@{\quad}r@{\quad}r@{\quad}r@{\quad\quad}rr@{}}
\toprule
& \multicolumn{4}{@{}c@{}}{Energy consumption (mWh for 10 minute-intervals)} &  \multicolumn{2}{@{\quad}c@{}}{Per Job} \\  \cmidrule(lr){2-5} \cmidrule(lr){6-7}
Device & Idle  & Upload  & Download  & Computation & Time (s) & Energy (mWh) \bigstrut \\
\midrule
Cloudlet & 5783.3  & 6038.3  & 5935.0  & 15706.7 & 1.5 & 39.5 \bigstrut \\
Google Nexus 9 & 433.3 & 476.7 & 556.7 & 1138.3  & 8.8 & 16.6 \bigstrut \\
Google Pixel 4 & 185.0  & 271.7  & 230.0 & 528.3 & 2.9 & 2.5 \bigstrut \\
Samsung Galaxy S7e  & 148.3 & 523.3 & 260.0  & 610.0 & 5.4 & 5.5 \bigstrut \\
Samsung Galaxy Tab S5e  & 316.7 & 521.7  & 385.0  & 741.7 & 4.0 & 5.0 \bigstrut \\
Xiaomi Mi 9T  & 151.7 & 250.0 & 178.3  & 535.0 & 3.2 & 2.9 \bigstrut \\
\bottomrule
\end{tabular}%
\end{table*}

The benchmark application runs on every device and fires object
detection jobs according to a Poisson process with a configurable job
inter-arrival time~$\lambda$, and relative deadlines~$d \le \lambda$.
Each job takes as input a randomly selected Ultra-HD image taken from
the UltraEye dataset~\cite{ultraeye}, and produces an objection detection
report of at most~$4$~KB. Each image has a pixel
resolution of $3840\times2160$ and an average size of 2.2~MB. All
images used were uploaded to the Android devices prior to benchmark execution
(as mentioned earlier, the cloudlet does not generate jobs in our experiments, it only executes them
on behalf of Android devices).

\section{Evaluation\label{s:eval}}
\vspace*{12pt}
We now present the detailed experiments we conducted and the
results we obtained. Their implications are also discussed.

\subsection*{Experiments}

Using the experimental setup described in the previous section, we
conducted three sets of experiments:
\begin{description}[style=unboxed,leftmargin=0cm]
\item [1.] We first measured the baseline behavior, in terms of energy
  and computation time, for the set of devices at hand, when executing
  the benchmark application in our setup without considering any type of
  offloading. The goal was to allow a relative comparison between
  devices given their heterogeneity.
\item [2.] We then considered offloading experiments for the benchmark
  application in a network formed only by the Android
  devices\trackchange{REVIEW}{
  \sout{.}, where each device acts both 
  as a job generator and worker.
  We compare the 
  use the {\rm LOCAL}, {\rm TMIN} and {\rm HYBRID} strategies for 
  job workloads with different values for mean job inter-arrival times 
  and job  deadlines.
  }
\item [3.] \trackchange{REVIEW}{\sout{Finally, }T}he previous experiments were repeated for the
  benchmark application this time using a network that also includes
  a cloudlet \trackchange{REVIEW}{worker.
  Again we consider {\rm TMIN} and {\rm HYBRID}
  strategies, but also the {\rm SERVER} strategy that offloads all jobs to the cloudlet
  server.
  }
\trackchange{REVIEW}{
 \item [4.] Finally, we consider again a network with
 mobile devices,  the effect of using the {\rm BALANCED} and ${\rm LF[f]}$ 
 strategies versus {\rm TMIN} and {\rm HYBRID}, plus a different choice of 
 configuration where jobs are generated and scheduled by a single external host
  and the mobile devices act only as workers, i.e., a Femtocloud-like
 configuration.
}
\end{description}

\subsection*{Baseline experiments}

For a baseline comparison between devices, we measured power
consumption and time/energy consumption during job execution for all
devices. We first ran scripts to measure (instantaneous) power
consumption when devices were idle, uploading data and downloading
data. Each script ran for 10 minutes and average power consumption
results were gathered from 3 script executions. For the
uploading/downloading power measurements the scripts continuously
executed plain file uploads/downloads to/from a random host in the
network.  For job computation behavior, we ran a script that issued
local object detection jobs continuously for 10 minutes, again for 3
rounds, and computed the average power consumption and job execution
time.

The results are listed, per device, in Table~\ref{tab:baseline}: power
consumption (in Watts, when devices are idle, uploading, downloading,
or computing); job execution time (seconds), and; energy consumption
(in milliwatt-hour, taking into account power consumption when
computing and the execution time per job).

Overall, the results clearly expose the heterogeneity of the devices
used in the experiments.  Looking at the power consumption results, it
is clear that computation is the major factor of increase in power
consumption: 2.3--4.1 times more energy is consumed than when a device
is idle, compared to just 1.1--3.5 times for uploading and 1.1--1.8
times for downloading. Compared to the Android devices, power
consumption numbers for the cloudlet are an order of magnitude higher
(approx.\ 10--40 times higher). Energy-wise, the two best-performing
devices while computing are Google Pixel~4 and Xiaomi Mi~9T. Samsung
Galaxy S7e is the most energy conservative device when in idle mode.

Regarding the results for execution time and energy consumption per
job, the cloudlet stands out again: it is both the most efficient
device in computation time, and the least efficient one in energy
consumption: jobs run 1.9--5.8 times faster than on the Android
devices but on the other hand consuming 2.4--15.6 times more
energy. Among the Android devices, and for both time and energy,
Google Pixel~4 is the most efficient device, followed by Xiaomi Mi~9T,
Samsung Galaxy Tab~S5e, and Samsung Galaxy~S7e, with Google Nexus~9
being the least efficient.

We note that the measures for energy consumption per job are more
relevant for our purposes (cf. Section~\ref{s:jay}) than those
for instantaneous power consumption.  Observe that Samsung Galaxy Tab
S5e is more energy-efficient (consumes $5.0$ mWh per job) than
Samsung Galaxy S7e (which consumes $5.5$ mWh per job, $10\%$ more),
even if instantaneous power consumption is higher during computation
($4.5$~W vs. $3.7$~W, $21\%$ higher).  The reason for this is that
the higher power consumption is compensated in a larger proportion by
faster job execution times in Samsung Galaxy Tab S5e ($4.0$~s
vs. $5.4$~s, $33\%$ faster).

As for the measured bandwidth during the duration of the experiment, we obtained
values averaging 110~Mbit/s for download on all mobile devices and for upload
we verified two distinct behaviors: Nexus 9, Pixel 4 and Samsung Galaxy S7e connected with a 300~Mbit/s connection averaging 210~Mbit/s speeds while Samsung Galaxy Tab S5e and Xiaomi Mi 9T connected
to the router with a 150~Mbit/s connection leading to an average upload speed of 119~Mbit/s.
As for the cloudlet, it was connected to our router via gigabit ethernet and we obtained
and average of 941~Mbit/s upload speed and 946~Mbit/s download.

\begin{figure*}[h!]
\includegraphics[width=\textwidth]{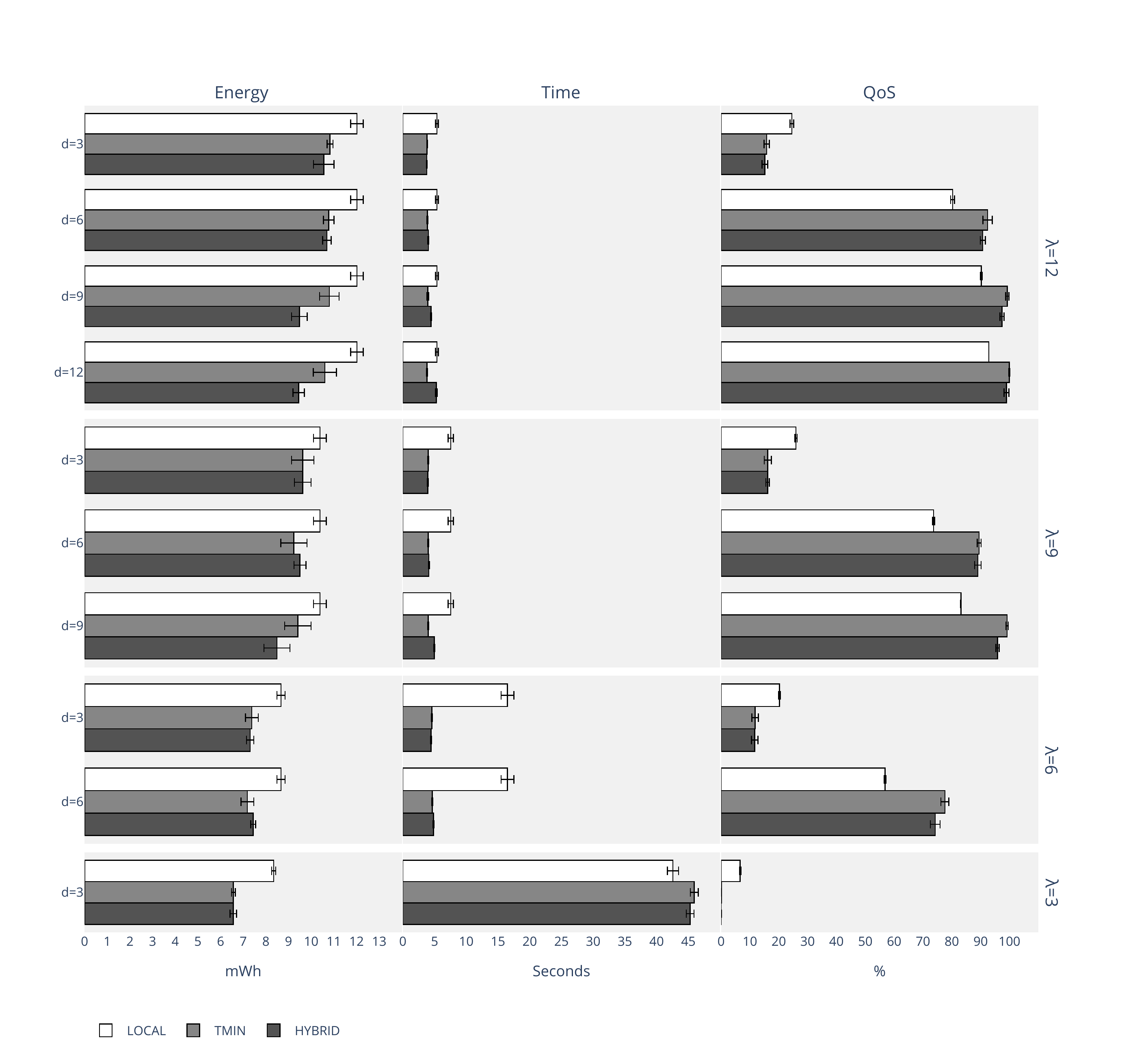}
\caption{Android devices scenario -- energy, time, and QoS.}
\label{fig:total_energy}
\end{figure*}

\begin{figure*}[h!]
\begin{subfigure}{0.45\textwidth}
\includegraphics[width=\textwidth]{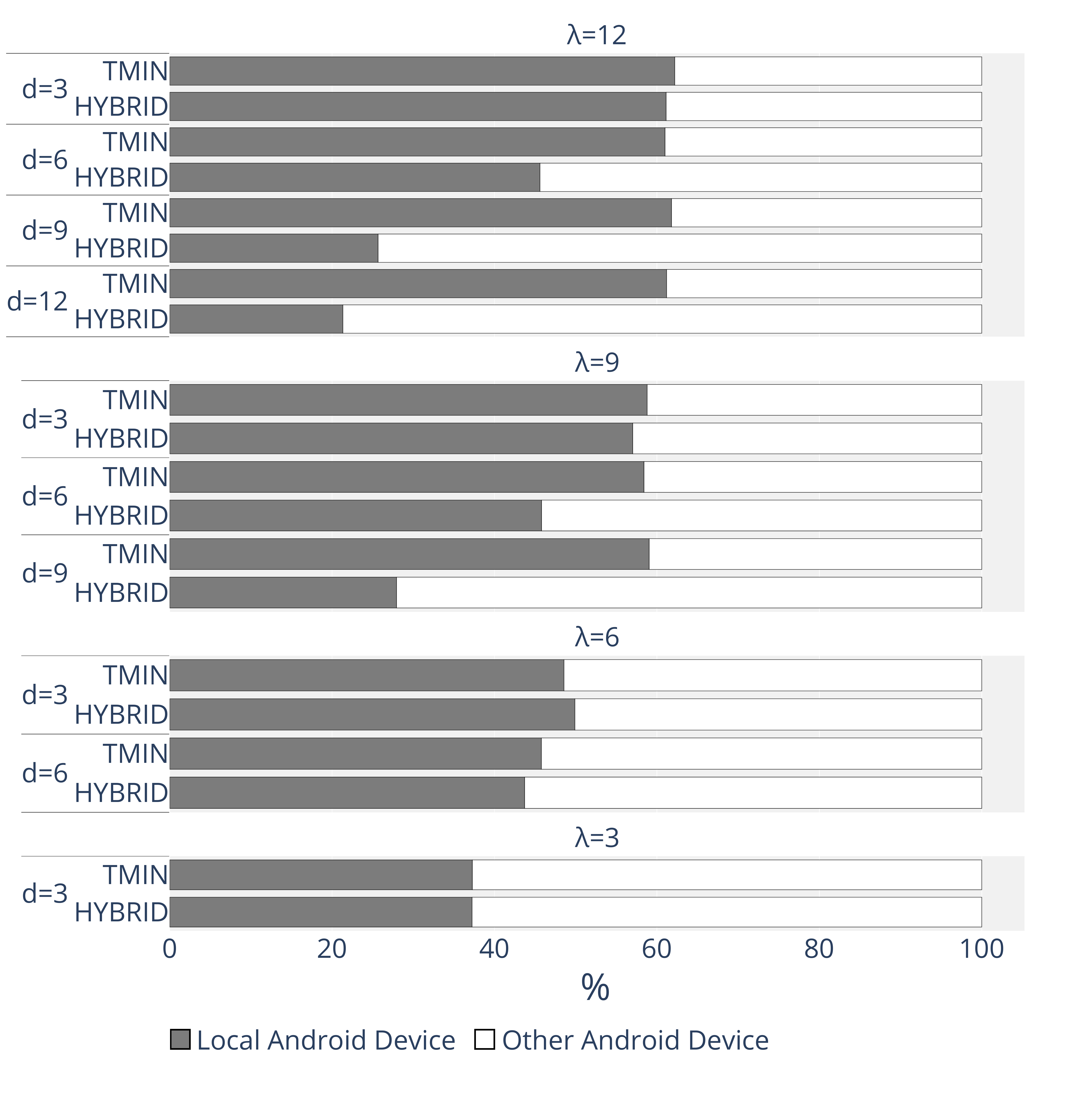}
\caption{Local vs. offloaded jobs.}
\label{fig:distribution:share}
\end{subfigure}
\begin{subfigure}{0.45\textwidth}
\includegraphics[width=\textwidth]{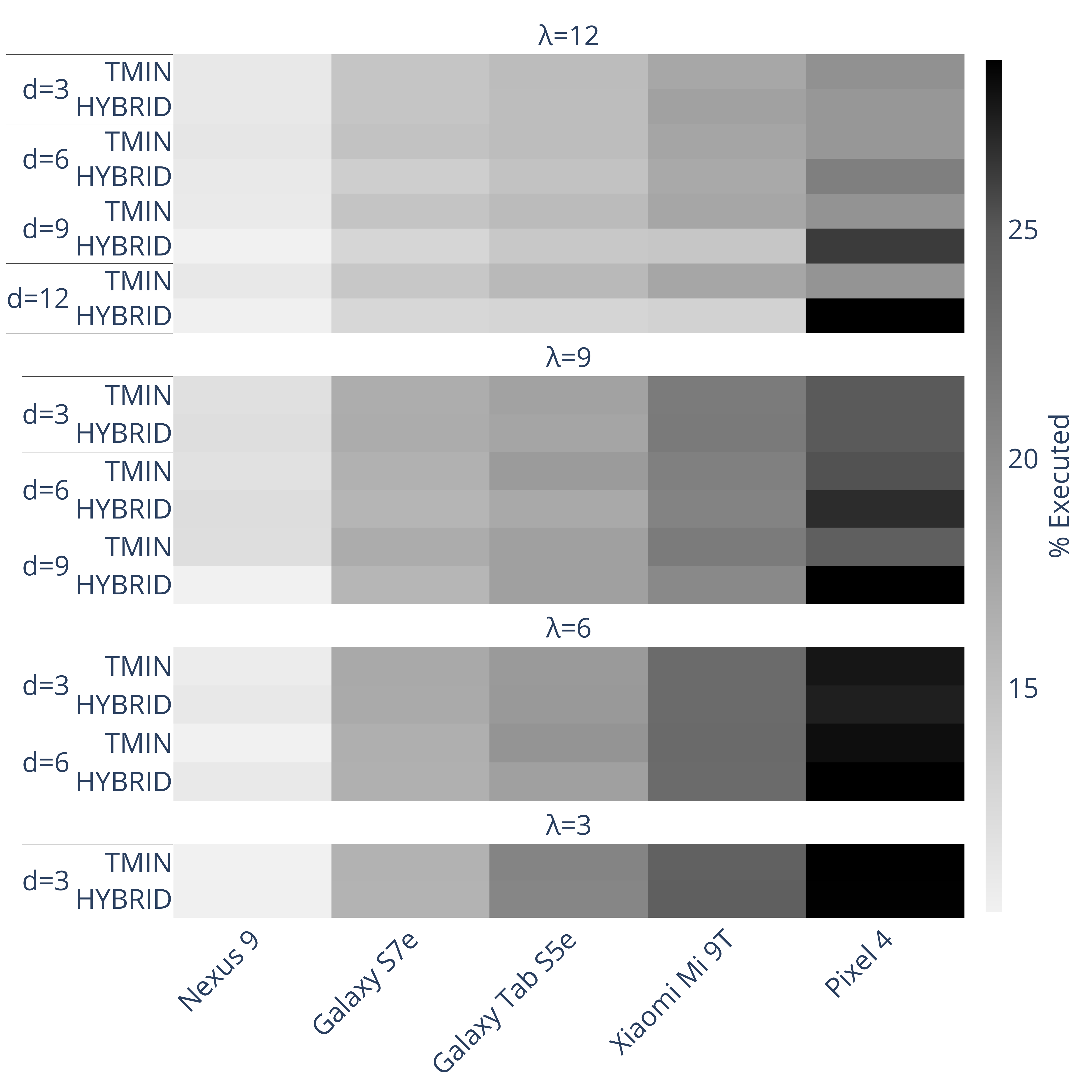}
\caption{Executed jobs per device.}
\label{fig:distribution:device}
\end{subfigure}
\caption{Android devices scenario -- job distribution.}
\label{fig:distribution}
\end{figure*}

\subsection*{Offloading among Android devices}

We considered a network formed by the Android devices, each running
the benchmark application generating jobs with mean inter-arrival
times for the governing Poisson process of $\lambda$ equal to $3$,
$6$, $9$ and $12$ seconds (which translates to $20$, $10$, $6.7$ and
$5$ jobs per minute respectively), and values of $d = 3, 6, 9, 12$ for
their relative deadlines up to the value of $\lambda$
(i.e., $d \leq \lambda$).

In conjunction, we considered three offloading strategies, presented
in Section~\ref{s:model}: {\rm LOCAL} (local execution only, no
offloading), {\rm TMIN} (offloads jobs strictly seeking to minimize
execution time) and {\rm HYBRID} (balances QoS constraints for
task deadlines with energy efficiency).  The benchmark was executed 6
times for each offloading strategy with the same job generation seed,
and each execution was configured to generate jobs for 10 minutes.


A first set of overall results for the experiment is presented in
Figure~\ref{fig:total_energy}.  We present plots for the energy
consumption and execution time per job (left and middle in the figure,
lower numbers are better), along with the corresponding
quality-of-service (QoS) that is expressed as the percentage of jobs
with a fulfilled deadline (right, higher numbers are better).  The
average values are plotted for each measure, along with the amplitude
of the $95\%$ gaussian confidence interval.  Note that, for each
configuration, the average energy consumption is obtained by measuring
the total energy consumption in all of the devices, including idle
time, divided by the number of jobs. Lower values of $\lambda$ imply
more jobs, hence the average energy consumption tends to decrease
with~$\lambda$ (conversely, idle time grows with~$\lambda$).

From the results, we can first observe that both TMIN and HYBRID
generally outperform LOCAL both in energy consumption and QoS. This shows that
offloading jobs pays off in both dimensions when compared to strictly
local execution of jobs. The exception to this pattern is observed when
the relative deadline has the tightest value, i.e., $d=3$, and only in
terms of QoS. In fact, the overall system becomes incapable of achieving
reasonable QoS in all configurations when at this point: always below $30\%$,
regardless of offloading strategy. In the more extreme case where
$\lambda = d = 3$, the QoS is below $10\%$ and there is an extremely
long execution time, due to the fact that jobs simply pile up in the
system. In contrast, the QoS is always higher than~$50\%$ for all
configurations with~$d>3$.

\begin{figure*}[h!]
\begin{subfigure}{0.45\textwidth}
\includegraphics[width=\textwidth]{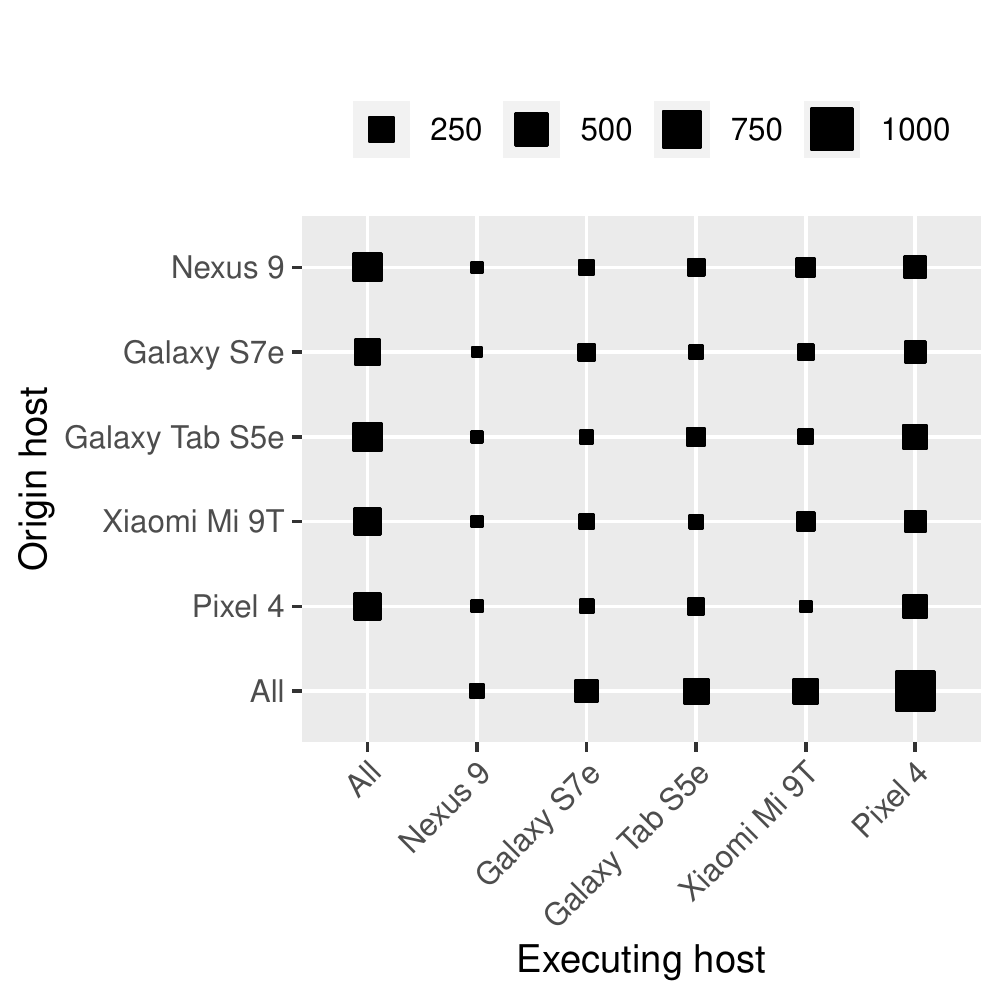}
\caption{HYBRID.}
\label{fig:exec:HYBRID}
\end{subfigure}
\begin{subfigure}{0.45\textwidth}
\includegraphics[width=\textwidth]{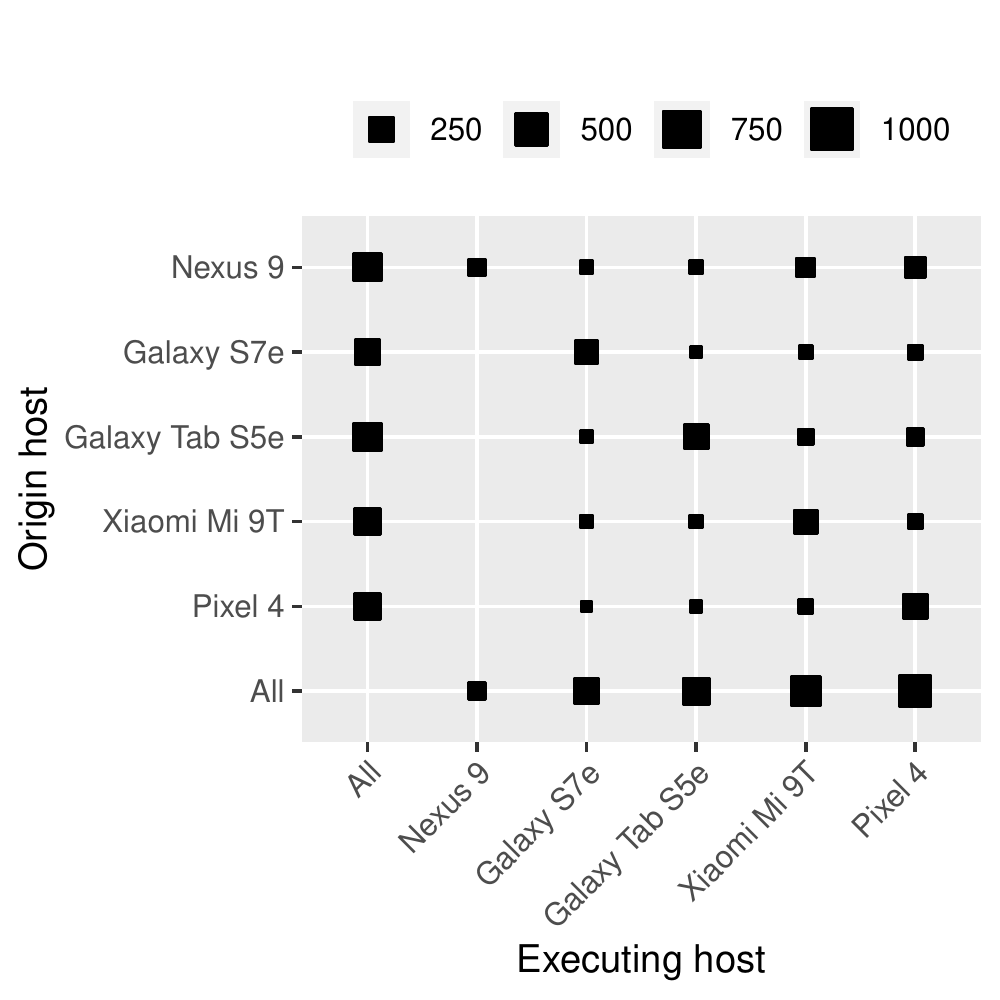}
\caption{TMIN.}
\label{fig:exec:TMIN}
\end{subfigure}
\caption{Android devices scenario -- flow of jobs for $\lambda=12$ and $d=9$.}
\label{fig:exec}
\end{figure*}

Comparing TMIN and HYBRID, the results are very similar for~$d=3$
and~$d=6$ in all respects (energy, time, and QoS).  Since these deadline
values are the most tight, the HYBRID strategy has less scope for
energy-efficient offloading choices and these tend to be similar to
the choices made by TMIN.  For $d>6$ there are noticeable differences
though, highlighting that gains in energy consumption can be attained
by the HYBRID strategy compared with TMIN at the cost of a slight
penalty in QoS.
The HYBRID strategy leads to a $10$--$20\%$ decrease in energy
consumption compared to TMIN, while the QoS service is only marginally
higher for TMIN, at most by~$5\%$. At the same time, the execution
time is slightly higher for HYBRID, given that the strategy does not
pick the the device estimated to run a job faster but, rather, the
most energy-efficient among those that are estimated to comply
with the job deadline. For example, when~$\lambda=12$ and~$d=9$ and
for HYBRID we observed: $13\%$ less energy consumption ($9.5$~mWh
compared to ${\sim}10.8$~mWh for TMIN); jobs taking $15\%$ longer
($4.5$ s vs. $3.9$ s), but; a QoS degradation of only $2\%$ ($97\%$
vs. $99\%$).

The behavior of TMIN and HYBRID is compared in more detail in
Figure~\ref{fig:distribution}, regarding the fraction of offloaded
jobs (\ref{fig:distribution:share}, left) and the fraction of jobs
executed per device (\ref{fig:distribution:device}, right).  These results
again illustrate that there is no significant difference between both
strategies for the tighter deadline of $d=3$. As the value of~$d$
grows, however, the offloaded job ratio tends to grow and be
significantly higher for the HYBRID strategy, whereas there are only
small variations for TMIN for each value of~$\lambda$.
When~$\lambda=12$ for instance, the offloading ratio increases
progressively in the case of HYBRID as $d$ grows from~${\sim}40\%$
when~$d=3$ up to~${\sim}80\%$ when~$d=12$, while for TMIN it is
${\sim}40\%$ for all values of~$d$.

If we look at the fraction of executed jobs per device
(Figure~\ref{fig:distribution:device}), we see that they are overall
in line with the baseline results, i.e., faster devices (which are
also more energy-efficient) execute more jobs.  For instance, Google
Nexus 9, the slowest device, executes the fewest jobs, while Google
Pixel 4, the fastest one, executes the most jobs. The total spread of
jobs is more uniform in the case of TMIN than with HYBRID, while
HYBRID tends to favor Google Pixel 4 significantly for $d\ge6$.  
\trackchange{Revised explanation, in line with reviewer comment.}{
These
aspects are illustrated in particular for the flow of jobs
when~$\lambda=12$ and~$d=9$, again comparing HYBRID (a) and TMIN (b),
in Figure~\ref{fig:exec}. 
The job distribution is noticeably more biased towards Google
Pixel 4 in the case of HYBRID: Google Pixel 4 executes $56\%$ of all
jobs for HYBRID compared to $33\%$ for TMIN.
TMIN offloads jobs more uniformly to the other
devices (note that the size of the squares grows logarithmically),
even if Nexus 9 only executes local jobs in the case of TMIN.
}

\begin{figure}[h!]
\includegraphics[width=0.9\columnwidth]{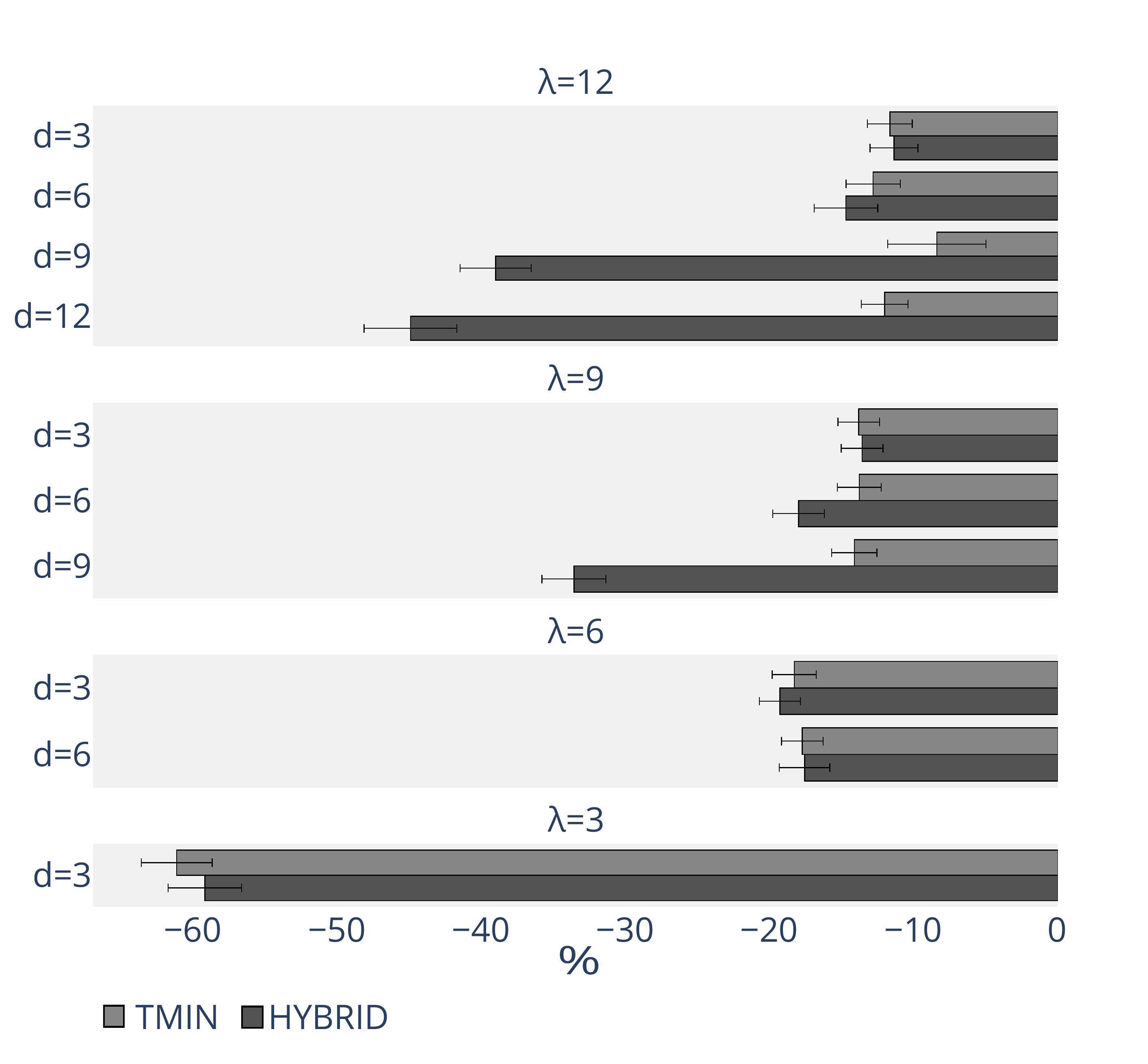}
\caption{Android devices scenario -- average error.}
\label{fig:avg_prediction_error}
\end{figure}

\begin{figure*}[h!]
\includegraphics[width=\textwidth]{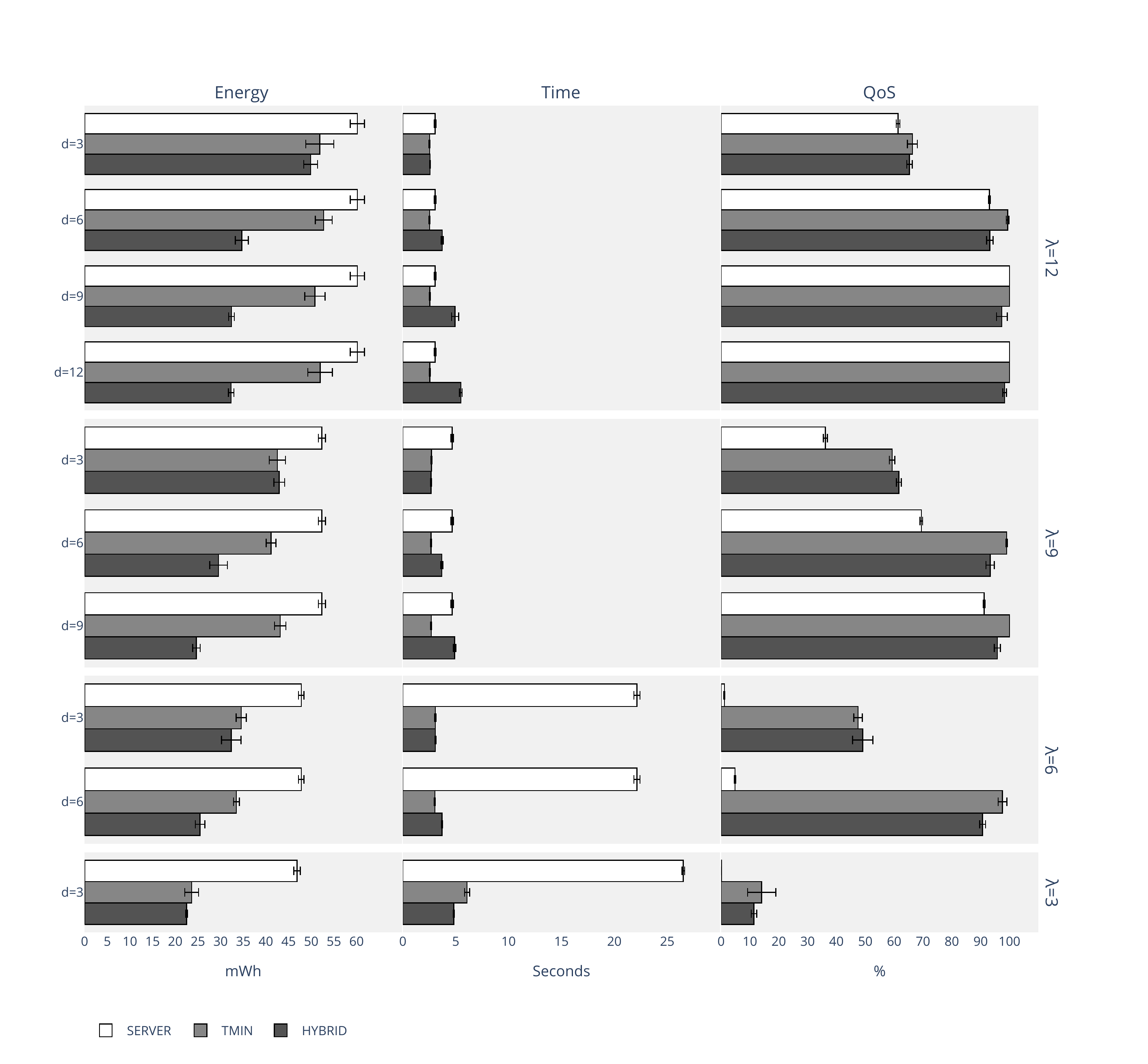}
\caption{Cloudlet scenario -- energy, time and QoS.}
\label{fig:total_energy_cloudlet}
\end{figure*}

\begin{figure*}[h!]
\begin{subfigure}{0.45\textwidth}
\includegraphics[width=\textwidth]{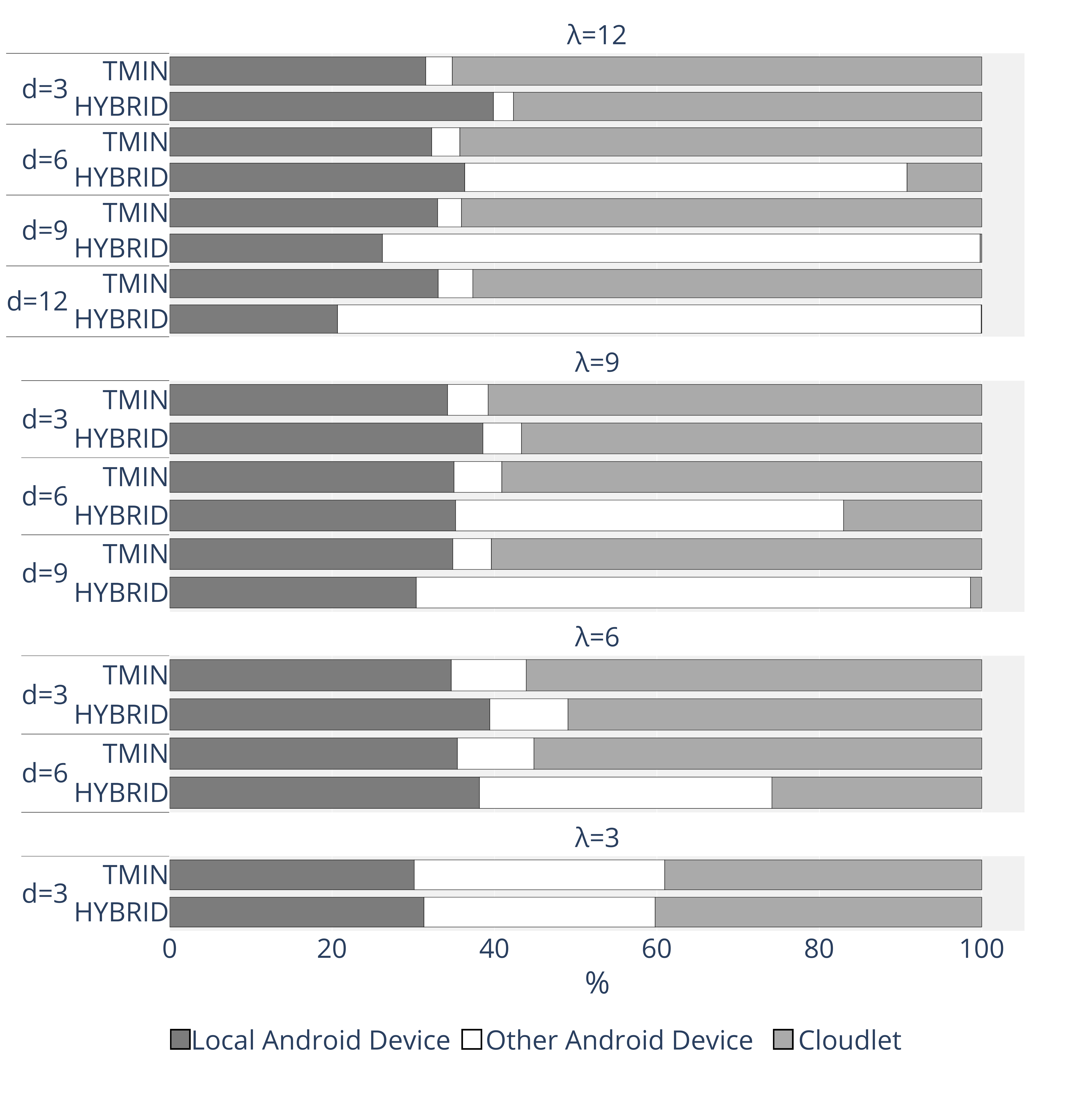}
\caption{Local, device, and cloudlet job share.}
\label{fig:execution_distribution_cloudlet}
\end{subfigure}
\begin{subfigure}{0.45\textwidth}
\includegraphics[width=\textwidth]{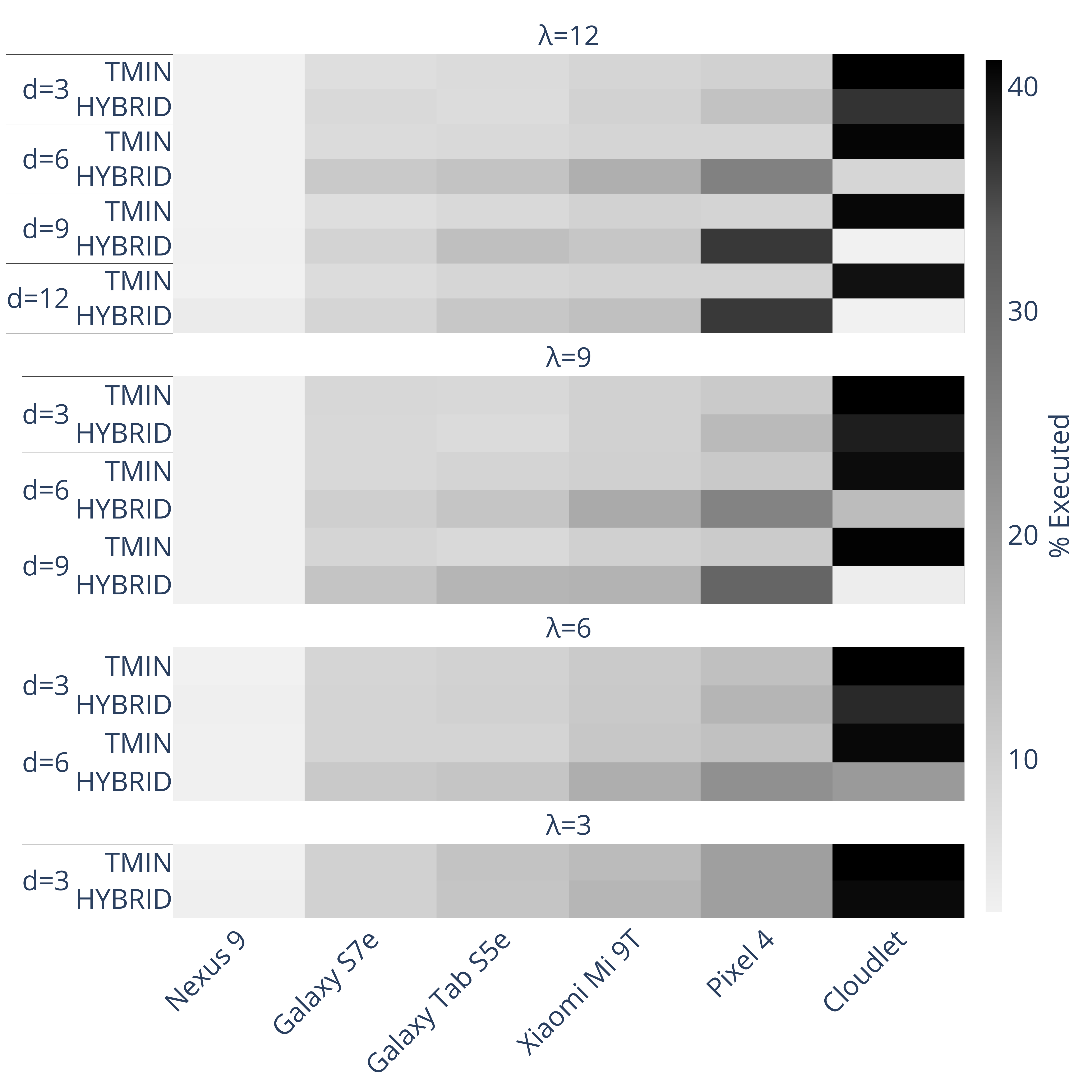}
\caption{Jobs per device.}
\label{fig:device_execution_distribution_cloudlet}
\end{subfigure}
\caption{Cloudlet scenario -- job distribution.\label{fig:dist_cloudlet}}
\end{figure*}

We finish our analysis by highlighting estimation errors by \Jay{}'s system profiler.
Figure~\ref{fig:avg_prediction_error} depicts the
average relative error in the estimated time for job execution,
calculated as the difference between estimated and real execution time,
expressed in percentage of the real execution time.
As shown, the values are on average negative, meaning that
the estimates tend to be pessimistic. In amplitude, they are less than~$20$\%
except for HYBRID when~$\lambda = 12,9$ and $d \geq 9$, and both strategies when $\lambda = 3$.
This is partly explained by the fact that
the estimate $\TE$ for execution time of a job at a \Jay{} instance
accounts for the current number of jobs including the current one,
but not the already spent executing the current job.
This behavior, which can be mitigated in future developments of the \Jay{} prototype, is amplified
in configurations where one the devices obtains a high share of jobs (Google Pixel 4 in the case
of HYBRID, for $\lambda = 12,9$ with $d \geq 9$). On the other hand, when the system has a high
load and is unable to cope (the case of $\lambda=3$), estimates also tend to be less reliable.

\subsection*{Extended scenario using cloudlet}

We now present results for an extension of the previous experiment
that introduces a cloudlet server. The cloudlet acts only as a \Jay{}
job executor, while job generation proceeds as before for the Android
devices. As before, the HYBRID and TMIN strategies were evaluated but
with the possibility of offloading from the devices to the
cloudlet. We consider, in addition, the {\rm SERVER} strategy that
uses the cloudlet as a standalone server that executes all jobs.  We
present measurements similar to the previous scenario and highlight
the impact of the cloudlet.

\begin{figure*}[h!]
\begin{subfigure}{0.45\textwidth}
\includegraphics[width=\textwidth]{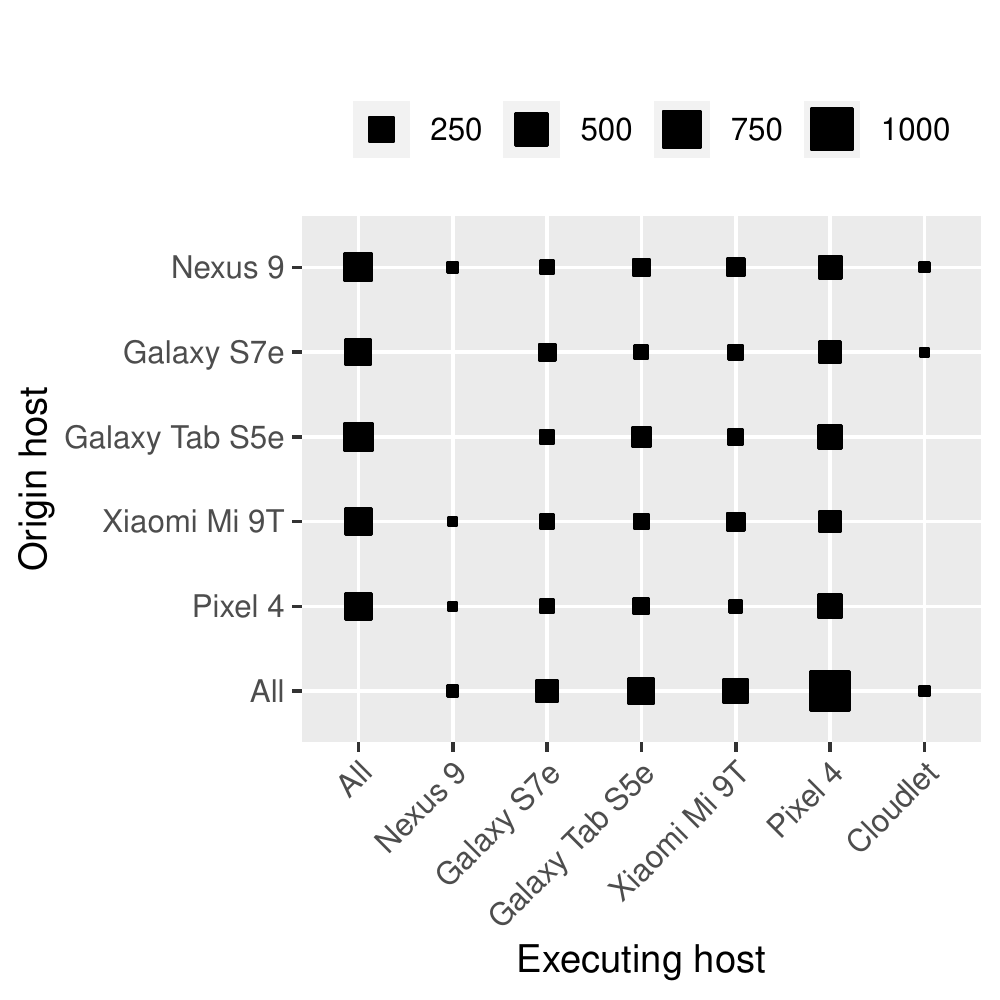}
\caption{HYBRID.}
\label{fig:cexec:HYBRID}
\end{subfigure}
\begin{subfigure}{0.45\textwidth}
\includegraphics[width=\textwidth]{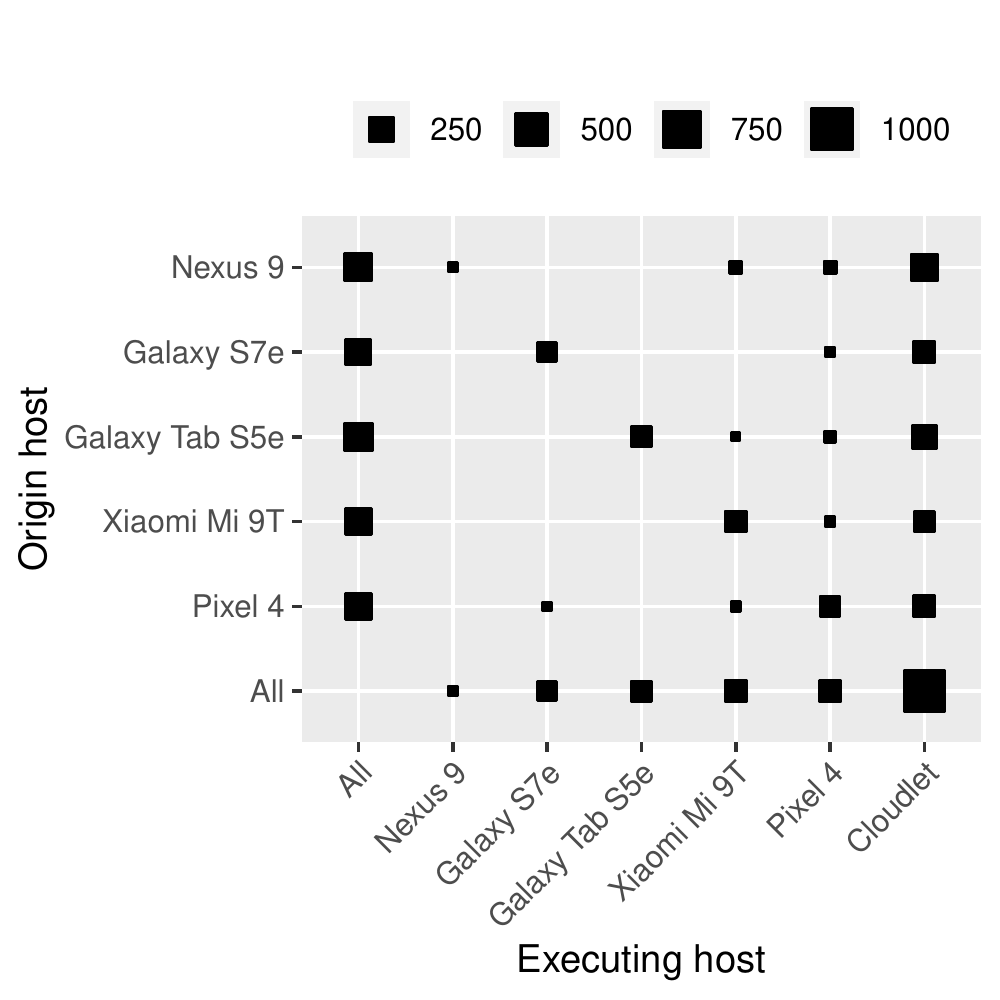}
\caption{TMIN.}
\label{fig:cexec:TMIN}
\end{subfigure}
\caption{Cloudlet scenario -- flow of jobs for $\lambda=12$ and $d=9$ (number of jobs).}
\label{fig:cexec}
\end{figure*}

In Figure~\ref{fig:total_energy_cloudlet} we provide plots for energy consumption, execution time
and QoS.  Compared to the results of the scenario without cloudlet
(cf. Figure~\ref{fig:total_energy}) an increase in energy consumption
as well as in QoS is noticeable for the TMIN and HYBRID
strategies. This would be expected, given that (in line with the
baseline results) the cloudlet is the most time-efficient device but
also the least energy-efficient one. The energy consumption is
significantly higher, something that will always be true even if the
cloudlet executes no jobs (in any case it will still actively consume
energy). For example, the lowest energy consumption value is $23$~mWh for HYBRID and
TMIN when $d=\lambda=3$, exceeding the value of the most energy-hungry
configuration of the previous scenario, $12$~mWh for $\lambda=d=12$ in
Figure~\ref{fig:total_energy}.  On the other hand, the cloudlet
improves QoS for HYBRID and TMIN significantly: it is now above $90\%$
for every configuration with $d\ge6$, and even $46\%$--$67\%$ for
$\lambda=12,9,6$ when $d=3$ in comparison to the $10$--$15\%$ observed
previously. QoS is very poor only, and again, in the
extreme~$\lambda = d = 3$ case.

Looking at the results for the SERVER strategy, they are generally
worse than those obtained for HYBRID and TMIN.  This is true for energy consumption
in all configurations, and also QoS except for configurations with
$\lambda=12$ where the SERVER strategy becomes competitive.  This
means that job execution/offloading by the Android devices pays off
compared to using the cloudlet alone, much like in the previous
scenario where it payed off when compared to using local job execution
only.

In this cloudlet scenario, energy consumption savings resulting from
the use of HYBRID vs. TMIN can be more pronounced.  In all
configurations, HYBRID consumes less energy than TMIN, and the savings
are noticeably more pronounced as~$d$ increases, e.g.,
for~$\lambda=12$, TMIN consumes just $4\%$  more
energy when $d=3$ but $60\%$ more when $d=12$. On the other hand, on
par with the decrease in energy consumption, HYBRID leads to
noticeably longer job execution times as $d$ grows, e.g., again for $\lambda=12$
HYBRID causes jobs to last from $2\%$ longer when $d=3$ up to $214\%$ when $d=12$.

The difference of behavior between HYBRID and TMIN is best understood
looking at the job distribution results in
Figure~\ref{fig:dist_cloudlet}, where we depict for all configurations
the fractions of: (\ref{fig:execution_distribution_cloudlet}) locally
executed jobs, jobs offloaded to Android devices, and jobs offloaded
to the cloudlet, and;
(\ref{fig:device_execution_distribution_cloudlet}) jobs per Android
device and cloudlet.  Besides the fact that HYBRID tends to have a
lower ratio of locally executed jobs, as in the previous Android
devices only scenario, the other major difference between HYBRID and
TMIN is that HYBRID tends to offload significantly less jobs to the
cloudlet than TMIN. Looking at the distribution per device, it is
clear that with HYBRID Google Pixel 4 is the device executing more
jobs, whereas TMIN privileges the cloudlet. In fact, in some
configurations, the fraction of jobs executed by the cloudlet can be
residual. The job flow for both strategies when $\lambda=12$ and $d=9$
is illustrated in Figure~\ref{fig:cexec}, and highlights this trend in
one of the more extreme cases: the cloudlet executes less than~$1\%$
of all jobs for HYBRID while Google Pixel 4 executes~$57\%$,
whereas for TMIN the fractions are $64\%$ for the cloudlet and $10\%$
for Google Pixel 4 (note that, as before, the size of the squares grows
logarithmically).

\begin{figure}[h!]
\includegraphics[width=0.9\columnwidth]{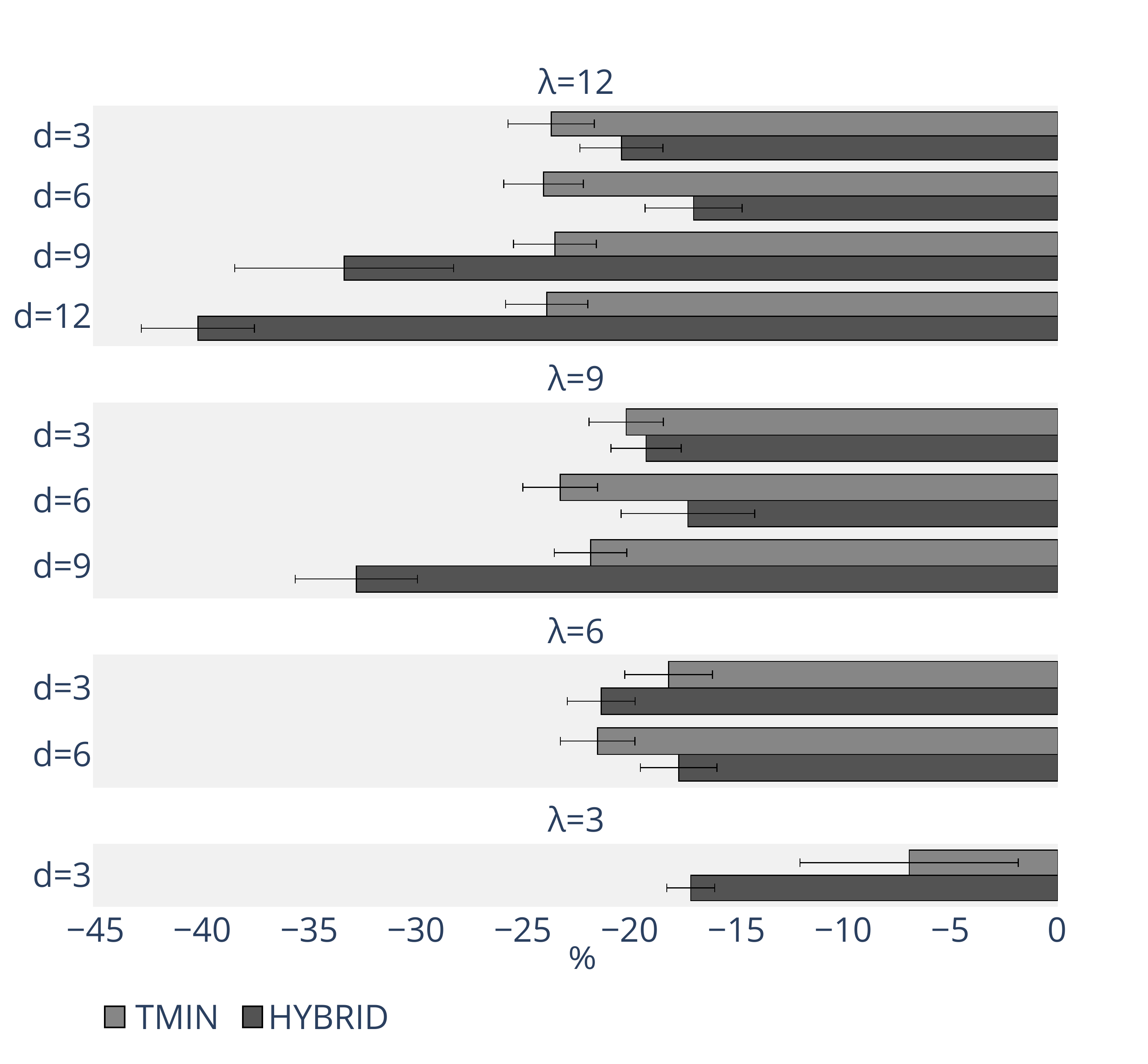}
\caption{Cloudlet scenario -- average error.}
\label{fig:avg_prediction_error_cloudlet}
\end{figure}

Estimation errors by \Jay{}'s system profiler are
presented in Figure~\ref{fig:avg_prediction_error_cloudlet} for the cloudlet scenario,
with similar trends to the Android devices' scenario (Figure~\ref{fig:avg_prediction_error}).
The main difference is
that estimation errors are not as high for the $\lambda=3$ case.
For this configuration, the overall system copes much better
with the high load scenario of in terms of job
execution times even if QoS is still low, and execution estimate errors tend to
be lower as a result.

\trackchange{Additional Scenarios}{

\subsection*{Additional Scenarios}

\begin{figure*}[h!]
\includegraphics[width=\textwidth]{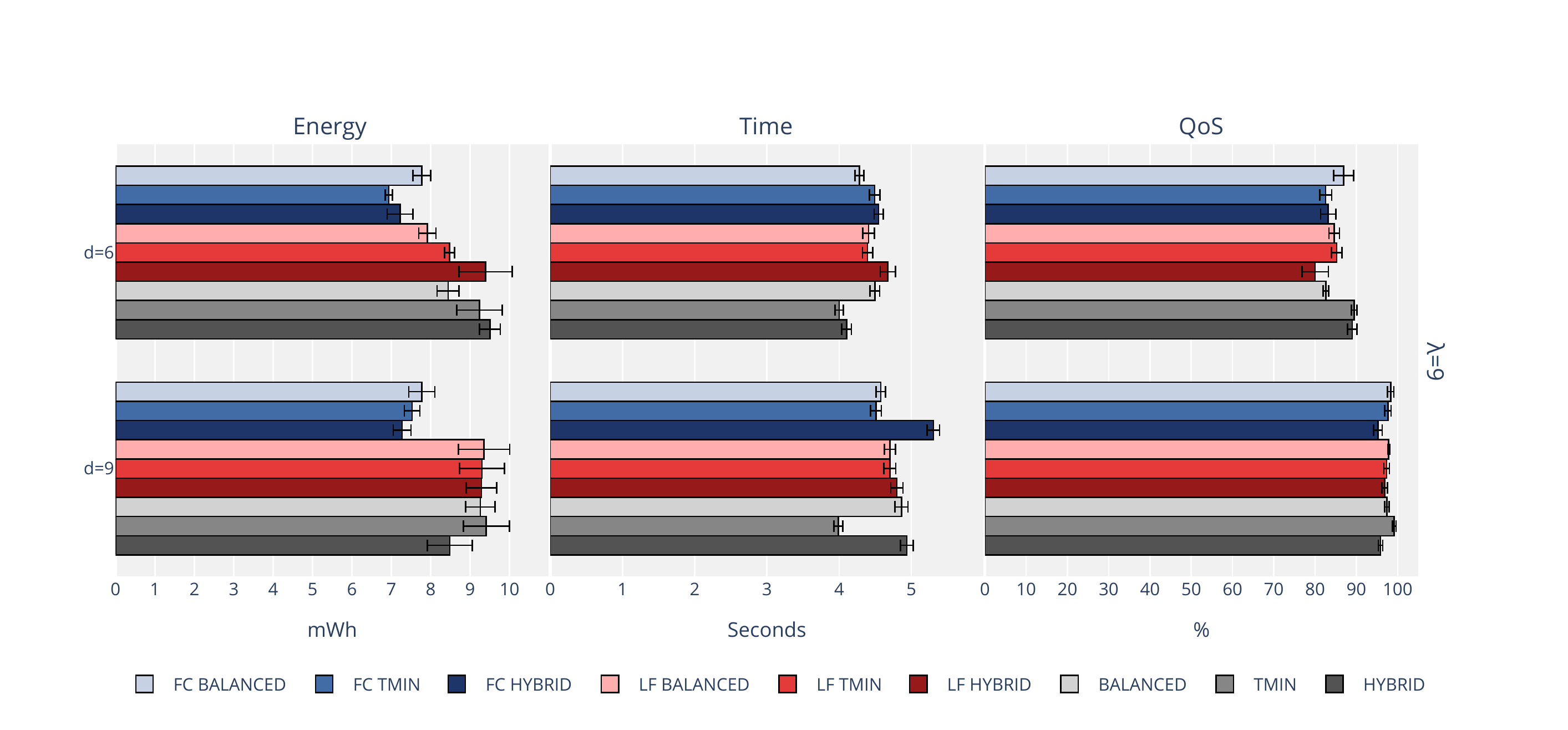}
\caption{Additional scenarios -- energy, time and QoS.}
\label{fig:total_energy_extra}
\end{figure*}

\begin{figure*}[h!]
\begin{subfigure}{0.45\textwidth}
\includegraphics[width=\textwidth]{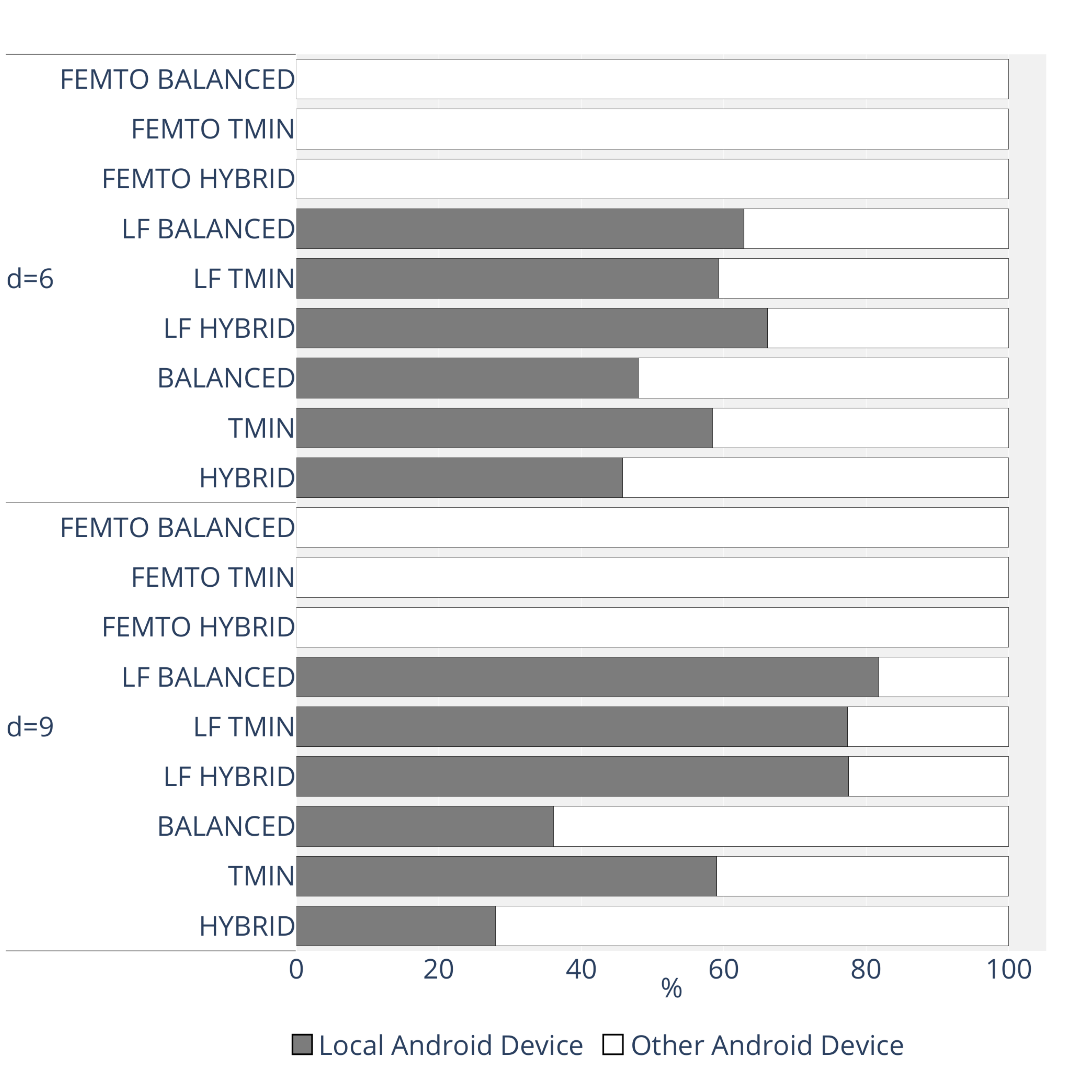}
\caption{Local and device job share.}
\label{fig:execution_distribution_extra}
\end{subfigure}
\begin{subfigure}{0.45\textwidth}
\includegraphics[width=\textwidth]{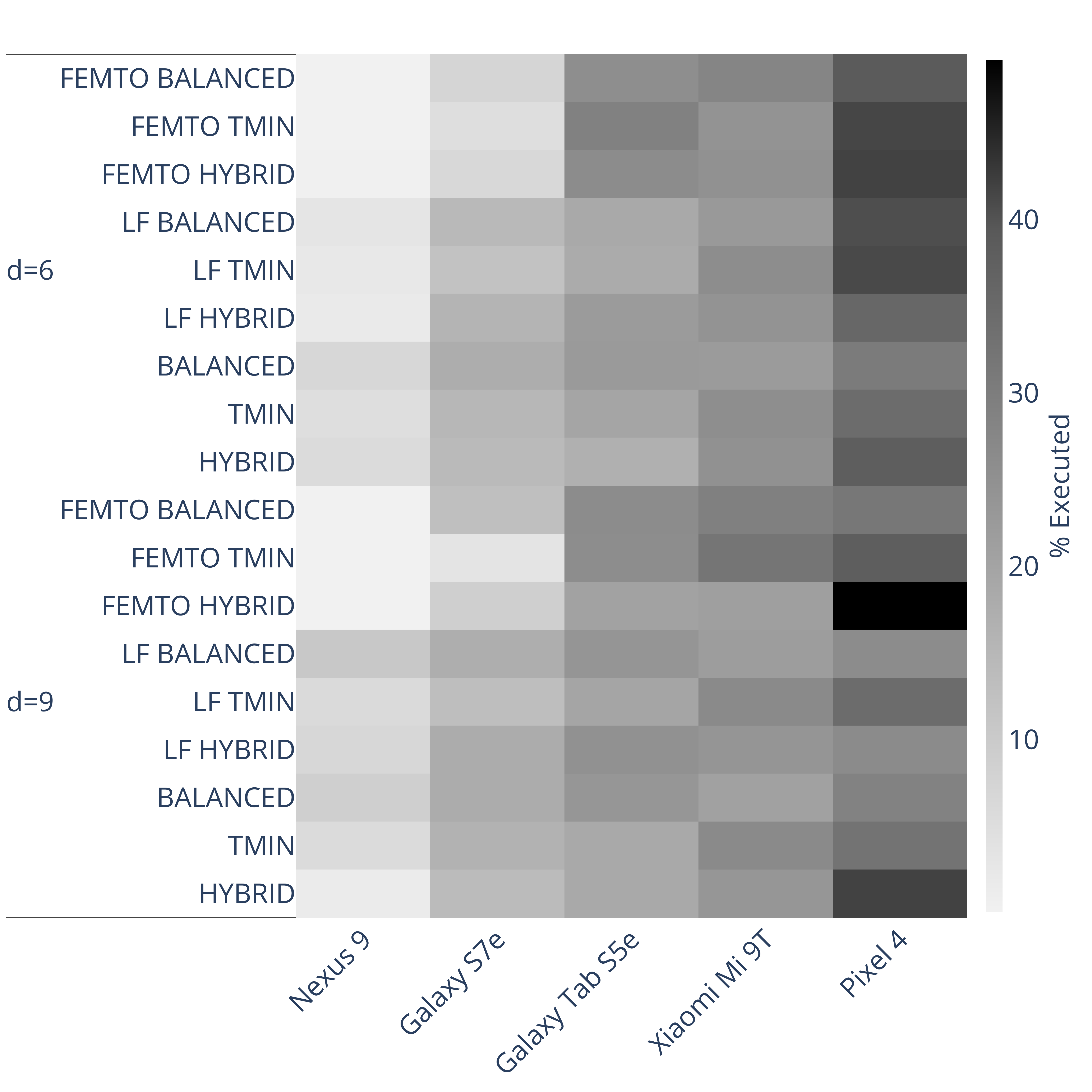}
\caption{Jobs per device.}
\label{fig:device_execution_distribution_extra}
\end{subfigure}
\caption{Additional scenarios -- job distribution.\label{fig:dist_extra}}
\end{figure*}

A final set of results is now presented, considering again a network
formed by mobile devices alone.  We consider the effect of having a
network with a Femtocloud configuration (FC), in which jobs are
generated and scheduled by a single external host and the mobile
devices act only as workers, in contrast to the mobile edge cloud
configuration (MEC) where all devices act as job generators and
workers.  Furthermore, we present results for additional offloading
strategies, {\rm BALANCED} and ${\rm LF[f]}$
(cf. Section~\ref{s:model}) that may potentially lead to different
compromises in terms of time, energy, and job distribution among
hosts.  {\rm BALANCED} applies both in the FC and MEC cases, whereas
${\rm LF[f]}$ (``local-first'') by definition only applies in the MEC
case (in the FC case, the external host does not act as worker, hence
it does not execute jobs locally).
 
Jobs were generated with the same methodology as in the previous
experiments the MEC configuration, but results were gathered only for
a job inter-arrival time of~$\lambda=9$ and deadlines~$d=6,9$.  In the
FC case, similar deadlines are considered, but the external host
generates jobs with a $\lambda=\frac{9}{5}$ inter-arrival time, so
that the overall workload is equivalent to the use of~$\lambda=9$ by
all~$5$ devices in the MEC configuration.  We empirically found these
workload parameterisations to be illustrative of the behavior of the
system for the strategies considered.

As in the previous experiments, the results are presented in terms of:
average energy consumption, average completion time and QoS
(Figure~\ref{fig:total_energy_extra}), and; job offloading rates and
job share per device (Figure~\ref{fig:dist_extra}).
  


\subsubsection*{Femtocloud setting}

Looking first at the FC results, the energy consumption values are
clearly the lowest, as shown in Figure~\ref{fig:total_energy_extra}
(top-left).  Compared to the MEC scenario, the energy consumption
values for TMIN and HYBRID are $14-20\%$ lower for $d=6$ and $25\%$
lower for~$d=9$. These gains, however, come at the cost of higher
execution times and lower QoS: for $d=6$ execution times $7-13\%$
higher, and the QoS is $4-7\%$ lower; for~$d=9$ the execution time are
$11-12\%$ higher but the QoS differences are small, lower than $2\%$
in absolute value.  Thus, the results are mixed, especially in the
case of $d=6$.

A priori, it would be expected that the centralised offloading
decisions to be more reliable in the FC configuration, since it is
free from the interference that arises from concurrent offloading
decisions by all devices in the MEC case.  However, in the MEC case
jobs can execute locally, e.g. for $d=6$ the share of local jobs is
$56\%$ for TMIN and $47\%$ for HYBRID, as depicted at the bottom in
Figure~\ref{fig:execution_distribution_extra}, and estimation errors
tend to be lower for locally executed jobs. In the FC case (by
definition) all jobs are offloaded leading to higher estimate
errors. These two factors influence the behavior in different
directions.

\subsubsection*{${\rm LF}[f]$ strategies}

By definition, ${\rm LF}[f]$ strategies try to execute as many jobs as
possible locally, resorting to offloading through strategy~$f$ only
when the local device is unable to cope with the deadline of a job.
Accordingly, as shown in Figure~\ref{fig:dist_extra}(a), the share of
locally executed jobs is significantly higher for ${\rm LF}[f]$ when
compared to $f$ in almost all cases, $15-49\%$ more, except for TMIN
when $d=6$ where the difference is negligible ($< 1\%$).

The results for ${\rm LF}[f]$ strategies are otherwise indicative of
energy/time/QoS trade-offs, as illustrated in
Figure~\ref{fig:total_energy_extra}. This happens especially
for~$d=6$.  In this case, when compared to TMIN, ${\rm LF}[{\rm
TMIN}]$ leads to a decrease of $9\%$ in energy consumption but, also,
an increase of $9\%$ in execution time and a decrease of $4\%$ in
QoS. This is expected as the base strategy, TMIN, seeks to minimize
execution time and thus it will tend to do better in this metric as
well as in QoS. Again for $d=6$, but using HYBRID as the base strategy
this time, ${\rm LF}[{\rm HYBRID}]$ degrades execution time and QoS by
even more, $12\%$ and~$9\%$ respectively, even if energy consumption
is roughly the same ($1\%$ difference between both). Given that most
energy-efficient devices tend to also be faster in our configuration,
the degradation of execution time and QoS is expected, as with TMIN,
but the difference in energy consumption is only noticeable for the
larger deadline value of $d=9$, where the energy consumption of ${\rm
LF}[{\rm HYBRID}]$ is $9\%$ higher.

\subsubsection*{The {\rm BALANCED} strategy}
Finally, the {\rm BALANCED} base strategy has the overall effect of
smoothing the load distribution among devices, as intended; recall
(from Section~\ref{s:model}) that the BALANCED strategy makes a random
choice among devices that are estimated to comply with a job's
deadline. Examining the numbers for the plot in
Figure~\ref{fig:dist_extra} (b), for configurations that employ TMIN
and HYBRID as a base or fallback strategy, the average shares of the
jobs for the Xiaomi Mi 9T and Pixel 4 devices combined (the two
devices that execute most jobs) are $64\%$ for $d=6$ and $63\%$ for
$d=9$. In comparison, for configurations that employ BALANCED as a
base or fallback strategy, the share of two devices is $3\%$ lower
($61\%$) for $d=6$ and, more noticeably, $10\%$ lower for $d=9$
($53\%$).  In more detail for $d=9$, the average individual share
grows for all of the 3 least used devices in the case of BALANCED:
from $3\%$ to $6\%$ for Nexus 9, from $12\%$ to $16\%$ for Galaxy S7e,
and from $21\%$ to $25\%$ for Galaxy Tab S5e. At the same time, the
average share in the case of BALANCED drops from $26\%$ to $24\%$ for
Xiaomi Mi 9T, and from $37\%$ to $29\%$ for Pixel 4.

Unlike the cases for the base strategies {\rm TMIN} and {\rm HYBRID},
the results for {\rm BALANCED} do not exhibit a clear trend with respect
to energy/time/QoS trade-offs (Figure~\ref{fig:total_energy_extra}).
All 6 executions per configuration use the same seed to guarantee a
repeatable job generation pattern. The particular job pattern may be
benefiting or hurting the behavior of {\rm BALANCED} in subtle ways
according to the configuration parameters, and the actual random
choice made by the {\rm BALANCED} strategy during execution. For
instance, for $d=6$ and the FC setting, the {\rm BALANCED} strategy
consumes $8-10\%$ more energy than TMIN and HYBRID, execution times
are $5-6\%$ faster, and QoS is $4\%$ higher. The trend is however
roughly symmetric for the MEC setting for instance: $9-11\%$ less
energy, $10-12\%$ slower execution times, and a $7\%$ lower QoS
value. 

}

\section{Conclusion} \label{s:conclusion}

In this paper we presented a model for soft-real time job offloading
over hybrid cloud topologies, along with offloading strategies that
try to optimize (either all or in part) execution time, total energy
consumption and fulfill QoS requirements in the form of job deadlines.
We instantiated the model in a software system, \Jay{}, and used it to
evaluate a variety of offloading strategies in clouds formed by mobile
devices and two-tier hybrid clouds formed by a network of mobile
devices and a cloudlet. \Jay{} is designed with adaptive scenarios in
mind. Offloading strategies are fed with the necessary runtime
information to perform time and energy-aware offloading on-the-fly,
Moreover, it employs a modular architecture that allows multi-tier
hybrid cloud topologies to be defined with customisable roles per tier
or device regarding job generation and execution.
The overall system flexibility was illustrated through experiments
using a benchmark application configured to spawn jobs with different
rates and different soft real-time deadlines, executed over different
cloud configurations and offloading strategies. The results of these
experiments show that offloading strategies sensitive to runtime
conditions can effectively and dynamically adjust their offloading
decisions to produce significant gains in execution time, energy
consumption and fulfillment of job deadlines.



For future work, we consider two key directions:
\begin{itemize}
\item Regarding application scenarios, we are particularly interested
  in articulating computation offloading with data-placement
  awareness, as in systems like Oregano~\cite{Sanches2020},
  \trackchange{Added reference to~\cite{huang2019multi}, 
  as suggested by Reviewer 1.}{our previous work on systems for data
  dissemination for hybrid edge clouds~\cite{Rodrigues2018,fmec20_ramble}, 
  which are particular instances of a class of systems that have multiple users,
  and employ multiple mobile devices, servers, and network tiers~\cite{huang2019multi}.
  }
  A challenge in these scenarios
  is that jobs may potentially require data stored at distinct hosts
  and/or tiers in the cloud, hence the interplay between computation
  and data offloading can potentially play a key role.  A different
  challenge is the possibility of high device churn and intermittent
  connectivity over heterogeneous communication links (WiFi, Bluetooth,
  4G/5G, etc), requiring offloading to proceed opportunistically, to
  be articulated with fault tolerance mechanisms (e.g., job
  checkpointing or replication), and the overall handling of a more
  dynamic environment regarding computational resources, network
  bandwidth, and energy consumption.
\item Regarding \Jay{} as a system, it can be extended in a number of
  ways to support a richer set of offloading strategies and job
  workloads.  Given its modular architecture, \Jay{} can easily
  accommodate for other multi-objective offloading strategies, of
  which the hybrid latency-energy offloading strategy is just an
  example, that account for additional aspects beyond execution time
  and energy consumption, e.g., the costs of using an infrastructure
  cloud or mobile device network traffic. 
  \trackchange{Added reference to TRACTOR~\cite{tractor}, as suggested by Reviewer 1.}{
  Moreover, even if \Jay{} is adaptive over a variety of hybrid cloud architectures,
  we believe that awareness of the cloud system used for offloading can lead
  to novel adaptive offloading strategies, e.g., as in the 
  TRACTOR algorithm~\cite{tractor} that accounts for aspects such as power consumption of network switches
  at the edge-cloud level for traffic and power-aware virtual machine placement.
  }
  \trackchange{}{Finally,} the system can also be
  improved for adaptivity in terms of resource awareness to cope with
  changeable cloud links due to mobility, and computational resources
  (e.g., GPUs could be used on the mobile devices by our deep learning
  benchmark).  Mobile applications also commonly exhibit features that
  would require our job model to richer, e.g., job precedences, job
  aggregation and their parallel execution, checkpointing to allow
  migration, etc.
\end{itemize}

\begin{backmatter}


\section*{Availability of data and materials}
\Jay{} is available as open-source software at \url{https://github.com/jqmmes/Jay/}.

\section*{Competing interests}
The authors declare that they have no competing interests.

\section*{Funding}
This work was partially funded by
project SafeCities (POCI-01-0247-FEDER-041435) from Fundo Europeu de
Desenvolvimento Regional (FEDER), through COMPETE 2020 and Portugal
2020.

\section*{Authors' contributions}

Joaquim Silva programmed \Jay{} and conducted the evaluation experiments.
All authors have participated in the conceptual design of the \Jay{} and 
associated experiments, data analysis, and manuscript writing. 

\section*{Acknowledgements}

The authors wish to thank the anonymous reviewers for the helpful
feedback.

\section*{Authors' information}

All authors are affiliated to the Department of Computer Science,
Faculty of Sciences, University of Porto (DCC/FCUP), and 
the Center for Research in Advanced Computing Systems at INESC TEC
(CRACS/INESC-TEC). 

Joaquim Silva is a PhD student in Computer Science at DCC/FCUP, 
Eduardo R. B. Marques is an assistant professor at DCC/FCUP,
Luís Lopes is an associate professor at DCC/FCUP,
and Fernando Silva is a full professor at DCC/FCUP.
All authors are researchers at CRACS/INESC TEC. 


%
%

%



\bibliographystyle{vancouver} 
\bibliography{refs}      


%
%

\end{backmatter}
\end{document}